# A systematic path to non-Markovian dynamics:
# New response pdf evolution equations under Gaussian coloured noise excitation

K.I. Mamis, G.A. Athanassoulis and Z.G. Kapelonis

konmamis@central.ntua.gr, mathan@central.ntua.gr, zkapel@central.ntua.gr

School of Naval Architecture & Marine Engineering, National Technical University of Athens
9 Iroon Polytechniou st., 15780 Zografos, GREECE

**Abstract.** Determining evolution equations governing the probability density function (pdf) of non-Markovian responses to random differential equations (RDEs) excited by coloured noise, is an important issue arising in various problems of stochastic dynamics, advanced statistical physics and uncertainty quantification of macroscopic systems. In the present work, such equations are derived for a scalar, nonlinear RDE under additive coloured Gaussian noise excitation, through the stochastic Liouville equation. The latter is an exact, yet non-closed equation, involving averages over the time history of the non-Markovian response. This nonlocality is treated by applying an extension of the Novikov-Furutsu theorem and a novel approximation, employing a stochastic Volterra-Taylor functional expansion around instantaneous response moments, leading to efficient, closed, approximate equations for the response pdf. These equations retain a tractable amount of nonlocality and nonlinearity, and they are valid in both the transient and long-time regimes for any correlation function of the excitation. Also, they include as special cases various existing relevant models, and generalize Hänggi's ansatz in a rational way. Numerical results for a bistable nonlinear RDE confirm the accuracy and the efficiency of the new equations. Extension to the multidimensional case (systems of RDEs) is feasible, yet laborious.

## TABLE OF CONTENTS

# 1. Introduction

The determination of the probabilistic structure of the response of a dynamical system to random excitations is an important question for many problems in statistical physics, material sciences, biomathematics, various engineering disciplines and elsewhere. In macroscopic stochastic dynamics it constitutes the basis of uncertainty quantification. For the special case of systems under white noise excitation, the answer is well-formulated; the evolution of the transition probability density function is governed by the Fokker-Planck-Kolmogorov (FPK) equation [1–3], [4] Sec. 5.6.6, [5] Ch. 5, [6] Sec. 6.3, [7] Sec. 3.18-9. In this case, the response is Markovian and, thus, its complete probabilistic structure is defined by means of the transition pdf. Unfortunately, this convenient description, via a single partial differential equation, is not applicable to problems where the random excitations are smoothly correlated (coloured) noises, and thus the responses are non-Markovian. The importance of coloured noise excitation in many advanced and real-life applications, and the theoretical complicacies it induces([1]), are discussed in many works, see e.g. [8], [9], Sec. IX.7, [10–16]. Despite the difficulties, various methods seeking to formulate response pdf evolution equations for systems under coloured noise excitation have been proposed and developed.

The most straightforward approach, popular in engineering applications, is the filtering approach [4] Sec. 5.10, [17,18]. In this case, the original system of random([2]) differential equations (RDEs) is augmented by a filter excited by white noise, whose output is utilized as an approximation of the given coloured excitation. Thus, the augmented system admits an exact FPK description. This approach is also called Markovianization by extension [19], or embedding in a Markovian process of higher dimensions [8], Sec. VI.A. Filtering is also the starting point of unified coloured noise approximation (UCNA) developed by Hänggi & Jung [20], [8] Sec. V.C. However important this approach may be, it has an inherent drawback; it leads to an inflation of the degrees of freedom in the FPK equation, especially when an accurate approximation of the excitation correlation is needed, in which case higher order filters are required.

An alternative approach is to try to derive pdf evolution equations analogous to the FPK equation, keeping only the natural degrees of freedom of the system, while taking into account *the given* coloured noise excitation. The main difficulty here, coming from the fact that "non-Markovian processes […] cannot be considered merely as corrections to the class of Markov processes" [21], lies in the emergence of terms dependent on the whole time history of the response, even for the one-time pdf evolution equation, as explained in detail subsequently. Therefore, closure techniques for the non-local term are needed, in order to obtain approximate, yet closed and solvable, forms of the evolution equations. Following Cetto et al. [22], we call these equations approximate generalized Fokker-Planck-Kolmogorov (genFPK) equations, since their white-noise excitation counterpart is the classical FPK equation.

---

([1]) In this case the infinite-dimensional character of the probabilistic description comes into play, preventing any straightforward construction of finite-dimensional probability evolution equations.
([2]) The term "random" is used herein, instead of "stochastic", to distinguish differential equations under coloured noise excitation from Ito stochastic differential equations, where the noise is assumed (usually tacitly) to be white. This terminological distinction has been also used by Ludwig Arnold [71].

Formulating genFPK equations was initiated in the 70's by the works of van Kampen [23], Fox [24] and Hänggi [25], developed further in the 80's, [22,26–29] (see also the seminal survey work of Hänggi and Jung [8]), and has been employed in many applications up to now. Examples of recent works applying genFPK equations to various disciplines are: [30] in energy harvesting, [31] in sensors design, [32] in laser technology, [33] in stochastic resonance, [16,34,35] in ecosystems, and [11,36,37] in medical science.

It should be observed, however, that the existing results are fragmented to some extent, with the connection between them still inconclusive; see for example [29], the discussion on Hänggi's ansatz in [28,38,39], as well as a recent comparison of the various existing genFPK equations in [34] Sec. 2.5. In the present work, the construction of genFPK equations is revisited, generalized and presented in a systematic and consistent way, by approximating the non-local terms by means of a novel methodology, that makes use of stochastic Volterra-Taylor functional series. A distinguishing feature of the proposed method is that the aforementioned Volterra-Taylor series are not considered around zero, but around appropriate instantaneous moments of the response, making it applicable to strongly nonlinear cases under high intensity noise excitation.

(a) *Nonlinear RDEs under coloured noise excitation and the corresponding stochastic Liouville equation*

In order to present the methodology in a clear way, we confine herein ourselves to the study of a simple prototype case, namely a scalar, non-linear, additively excited RDE:

$$\dot{X}(t;\theta) = h(X(t;\theta)) + \kappa \Xi(t;\theta), \qquad X(t_0;\theta) = X_0(\theta), \qquad (1a,b)$$

where $\theta$ is the stochastic argument, an overdot denotes time differentiation, $h(x)$ is a deterministic continuous function modelling the nonlinearities (restoring term), and $\kappa$ is a constant. Initial value $X_0(\theta)$ and excitation $\Xi(t;\theta)$ are considered correlated and jointly Gaussian with (non-zero) mean values $m_{X_0}$, $m_{\Xi(\cdot)}(t)$, autocovariances $C_{X_0 X_0}$, $C_{\Xi(\cdot)\Xi(\cdot)}(t,s)$ and cross-covariance $C_{X_0 \Xi(\cdot)}(t)$. Note that, in most of the existing works deriving genFPK equations [8,22,26,28,29], the initial value-excitation correlation is not taken into consideration, even though its importance has been recognized [38]. The influence of the correlation between initial value and excitation has been tackled by Roerdink [40,41] albeit for linear equations and mainly at the level of moments.

Since most of the existing genFPK equations have been derived for scalar RDEs, the choice of Eq. (1) as a prototype serves also the purpose of comparing our results to the existing ones. However, it is important to note that the whole methodology presented in this paper can be generalized to $N$–dimensional systems of nonlinear RDEs. First results in this direction are presented in a recent paper of ours [42].

The starting point of our analysis is the same as in many previous works [11,22,25,26,28,43,44]. The response pdf is represented as the average of a *random delta function*:

$$f_{X(t)}(x) = \mathbb{E}^{\theta}[\delta(x - X(t;\theta))]. \qquad (2)$$

Then, by differentiating both sides of the above equation with respect to time and using the identity $\partial \delta(x - X(t;\theta))/\partial t = -\dot{X}(t;\theta)\, \partial \delta(x - X(t;\theta))/\partial x$ and Eq. (1), we obtain the *stochastic Liouville equation*([3]) (SLE) corresponding to RDE (1), which reads as

$$\frac{\partial f_{X(t)}(x)}{\partial t} + \frac{\partial}{\partial x}\left(h(x) f_{X(t)}(x)\right) = -\kappa \frac{\partial}{\partial x}\left(\mathbb{E}^{\theta}\left[\Xi(t;\theta)\, \delta(x - X(t;\theta))\right]\right). \tag{3}$$

In Eq. (3) and subsequently, $\mathbb{E}^{\theta}[\bullet]$ denotes the ensemble average operator with respect to the response-excitation probability measure $\mathbf{P}_{X(\cdot)\Xi(\cdot)}$. A detailed derivation of Eq. (3) is given in Appendix A in the electronic supplementary material (ESM). This approach is also popular in the theory of turbulence, where is called the *pdf method* [45,46]. Herein, we propose and employ the term *delta projection method*, motivated by Eq. (2). Note that SLE (3) is alternatively derived by employing van Kampen's lemma [9,23,47]. A comparison between our derivation and van Kampen's derivation is presented in Appendix A, Sec. A(c), in ESM.

SLE (3) is exact, yet not closed, due to the presence of the term

$$\mathcal{N}_{\Xi X} = \mathbb{E}^{\theta}\left[\Xi(t;\theta)\, \delta(x - X(t;\theta))\right]. \tag{4}$$

There are two ways to proceed further with the term $\mathcal{N}_{\Xi X}$; either to change our considerations and turn to the study of the joint response-excitation pdf $f_{X(t)\Xi(t)}(x, u)$, a way of work proposed in [10,11,13], or to eliminate the explicit dependence of $\mathcal{N}_{\Xi X}$ on the stochastic excitation $\Xi(t;\theta)$, in which case the term $\mathcal{N}_{\Xi X}$ becomes non-local in time.

In the present work we follow the second approach. In this conjunction, the response $X(t;\theta)$ is seen, through the solution of RDE (1), as a function-functional (FF$\ell$) on the initial value $X_0(\theta)$ and the excitation $\Xi(\cdot;\theta)$ over the time interval $[t_0, t]$ (from the initial time $t_0$ to the current time $t$). The notation $X(t;\theta) = X[X_0(\theta); \Xi(\bullet|_{t_0}^{t};\theta)]$ is used subsequently, whenever it is important to declare the dependence of $X(t;\theta)$ on $X_0(\theta)$ and $\Xi(\bullet;\theta)$. The above discussion makes clear that $\mathcal{N}_{\Xi X}$ is a non-local term, depending *on the whole history of the excitation*. In a second step (Sec. 2), by an application of a new, extended form of the Novikov-Furutsu (NF) theorem, $\mathcal{N}_{\Xi X}$ is equivalently written as a functional *on the whole history of the response*, which is an ample manifestation of the non-Markovian character of the response. The said extension of the NF Theorem, recently derived by the same authors [48], is able to treat correlations between initial value and excitation, as well as non-zero mean excitation. Using the present approach, the classical FPK equation for RDE (1) under white noise excitation is easily rederived, as well as various existing genFPK equations corresponding to coloured noise excitation.

---

([3]) The term "stochastic Liouville equation" was introduced by Kubo in [72].

**(b)** *New, closed probability evolution equations for non Markovian response*

In Sec. 3, a novel approximation that employs Volterra-Taylor series around moments of the response, called the *Volterra adjustable decoupling approximation* (VADA), is derived. By applying this approximation at the nonlocal term of the SLE, we obtain a family of VADA genFPK equations, having the general form

$$\frac{\partial f_{X(t)}(x)}{\partial t} + \frac{\partial}{\partial x}\left[\left(h(x) + \kappa\, m_{\Xi(\cdot)}(t)\right) f_{X(t)}(x)\right] = \frac{\partial^2}{\partial x^2}\left[\mathcal{B}[f_X\,;x,t]\, f_{X(t)}(x)\right]. \quad (5)$$

The diffusion coefficient $\mathcal{B}[f_X\,;x,t]$, defined by Eq. (44), apart from being a function of state and time variables, is also a functional on the unknown response pdf, reflecting the non-Markovian character of the response. Thus, unlike other existing approaches, VADA technique retains an amount of nonlocality and nonlinearity of the original SLE (3), albeit of tractable character. Eq. (5) falls into the category of *nonlinear and nonlocal evolution equations*, usually called *nonlinear FPK equations* in the literature [49].

The new genFPK Eq. (5) exhibits the following plausible features:

- It is valid in both transient and long-time, steady-state regimes,
- It is valid for large correlation times and large noise intensities of the excitation,
- It yields the exact Gaussian solution pdf in the case of linear RDE,
- It applies to general Gaussian excitation, characterized by any correlation function,
- It applies to non-zero mean excitation, also correlated with the initial value.

Despite their fundamental nature, the above features are not simultaneously present in the existing genFPK equations. First numerical results, for a case of a bistable RDE, presented in Sec. 4, confirm the validity and the accuracy of VADA genFPK equations, by comparisons with results obtained from Monte Carlo simulations. In the last Sec. 5, some concluding remarks and a discussion on possible generalizations are presented.

## 2. Transformed stochastic Liouville equation and some straightforward applications

As proved in a recent work by the same authors [48], the extended NF theorem for a generic $\mathrm{FF}\ell$ $\mathcal{F} = \mathcal{F}[X_0(\theta)\,;\, \Xi(\bullet|_{t_0}^{t}\,;\theta)]$, whose arguments are jointly Gaussian, reads as follows:

$$\mathbb{E}^\theta\left[\Xi(t\,;\theta)\,\mathcal{F}[X_0(\theta)\,;\Xi(\bullet|_{t_0}^{t}\,;\theta)]\right] = m_{\Xi(\cdot)}(t)\,\mathbb{E}^\theta\left[\mathcal{F}[X_0(\theta)\,;\Xi(\bullet|_{t_0}^{t}\,;\theta)]\right] +$$

$$+ C_{X_0\Xi(\cdot)}(t)\,\mathbb{E}^\theta\left[\frac{\partial \mathcal{F}[X_0(\theta)\,;\Xi(\bullet|_{t_0}^{t}\,;\theta)]}{\partial X_0(\theta)}\right] + \int_{t_0}^{t} C_{\Xi(\cdot)\Xi(\cdot)}(t,s)\,\mathbb{E}^\theta\left[\frac{\delta \mathcal{F}[X_0(\theta)\,;\Xi(\bullet|_{t_0}^{t}\,;\theta)]}{\delta \Xi(s\,;\theta)}\right] ds,$$

(6)

where $\delta \bullet / \delta \Xi(s\,;\theta)$ denotes the Volterra functional derivative of $\mathcal{F}$ with respect to $\Xi(s\,;\theta)$. Since $\delta(x - X(t\,;\theta)) = \delta\!\left(x - X[X_0(\theta)\,;\Xi(\bullet|_{t_0}^{t}\,;\theta)]\right)$, the random delta function can be consid-ered as a FF$\ell$ like $\mathcal{F}$. Thus, we are able to apply the NF theorem (6) to the non-local term $\mathcal{N}_{\Xi X}$, Eq. (4), transforming SLE (3) into

$$\frac{\partial f_{X(t)}(x)}{\partial t} + \frac{\partial}{\partial x}\left[\left(h(x) + \kappa m_{\Xi(\cdot)}(t)\right) f_{X(t)}(x)\right] =$$

$$= \kappa\, C_{X_0 \Xi(\cdot)}(t)\, \frac{\partial^2}{\partial x^2}\, \mathbb{E}^{\theta}\!\left[\delta(x - X(t;\theta))\, \frac{\partial X[X_0(\theta);\Xi(\cdot|_{t_0}^{t};\theta)]}{\partial X_0(\theta)}\right] + \quad (7)$$

$$+ \kappa\, \frac{\partial^2}{\partial x^2} \int_{t_0}^{t} C_{\Xi(\cdot)\Xi(\cdot)}(t,s)\, \mathbb{E}^{\theta}\!\left[\delta(x - X(t;\theta))\, \frac{\delta X[X_0(\theta);\Xi(\cdot|_{t_0}^{t};\theta)]}{\delta \Xi(s;\theta)}\right] ds.$$

**Remark 2.1:** While the problem of correlation between initial value and excitation has been studied before by Roerdink [40,41] (for linear RDEs under general, possibly non Gaussian excitation), the use of the aforementioned extension of NF theorem poses a significant advantage. It explicitly incorporates the effect of initial value correlation into the transformed SLE (7). Thus, all genFPK equations derived from SLE (7) will inherit this effect in a straightforward way.

By comparing the transformed SLE (7) to its previous form, Eq. (3), we observe that the use of the NF theorem results in: **i)** an augmented drift term, which can be identified as the right-hand side of RDE (1) with excitation replaced by its mean value, **ii)** the appearance of second order $x$–derivatives in the right-hand side of the equation, and **iii)** the appearance of the averages of the of random delta function multiplied by the variational derivatives of the response with respect to initial value and excitation.

The variational derivatives appearing in Eq. (7),

$$V_{X_0}(t;\theta) = \frac{\partial X[X_0(\theta);\Xi(\cdot|_{t_0}^{t};\theta)]}{\partial X_0(\theta)}, \quad V_{\Xi(s)}(t;\theta) = \frac{\delta X[X_0(\theta);\Xi(\cdot|_{t_0}^{t};\theta)]}{\delta \Xi(s;\theta)}, \quad (8a,b)$$

can be calculated by formulating and solving the corresponding variational equations. The latter are formally derived from RDE (1), by applying the differential operators $\partial \bullet / \partial X_0(\theta)$ and $\delta \bullet / \delta \Xi(s;\theta)$, respectively [50] Sec. 2.7, [51] Ch. II Sec. 9, [52] Sec. 2.10:

$$\dot{V}_{X_0}(t;\theta) = h'(X(t;\theta))\, V_{X_0}(t;\theta), \qquad V_{X_0}(t_0;\theta) = 1, \quad t > t_0, \quad (9a,b)$$

$$\dot{V}_{\Xi(s)}(t;\theta) = h'(X(t;\theta))\, V_{\Xi(s)}(t;\theta), \qquad V_{\Xi(s)}(s;\theta) = \kappa, \quad t > s. \quad (10a,b)$$

Note that, the initial value problem (10) is defined for $t > s$ since, by causality, any perturbation $\delta \Xi(s;\theta)$, acting at time $s$, cannot result in a perturbation $\delta X(t;\theta)$ for $t < s$; thus, $V_{\Xi(s)}(t;\theta) = 0$ for $t < s$. Since variational Eqs. (9), (10) are linear ordinary differential equations (ODEs) with respect to $t$, their solutions are explicitly obtained:

$$V_{X_0}(t;\theta) = \exp\!\left(\int_{t_0}^{t} h'(X(u;\theta))\, du\right), \quad (11a)$$

$$V_{\Xi(s)}(t;\theta) = \kappa \exp\left(\int_s^t h'(X(u;\theta))\, du\right), \tag{11b}$$

where the prime denotes the derivative of function $h(x)$ with respect to its argument. To simplify the notation, we set

$$\mathcal{I}_{h'}[X(\bullet|_s^t;\theta)] = \int_s^t h'(X(u;\theta))\, du. \tag{12}$$

Substituting now the solutions (11) into Eq. (7), and using the notation (12), we obtain the following new (final) form of the SLE

$$\begin{aligned}\frac{\partial f_{X(t)}(x)}{\partial t} &+ \frac{\partial}{\partial x}\left[\left(h(x) + \kappa m_{\Xi(\bullet)}(t)\right) f_{X(t)}(x)\right] = \\ &= \kappa\, C_{X_0 \Xi(\bullet)}(t)\, \frac{\partial^2}{\partial x^2}\, \mathbb{E}^\theta\!\left[\delta(x - X(t;\theta))\, \exp\!\left(\mathcal{I}_{h'}[X(\bullet|_{t_0}^t;\theta)]\right)\right] + \\ &+ \kappa^2\, \frac{\partial^2}{\partial x^2}\, \mathbb{E}^\theta\!\left[\delta(x - X(t;\theta)) \int_{t_0}^t C_{\Xi(\bullet)\Xi(\bullet)}(t,s)\, \exp\!\left(\mathcal{I}_{h'}[X(\bullet|_s^t;\theta)]\right) ds\right]. \end{aligned} \tag{13}$$

SLE (13) contains two non-local terms in its right-hand side, carrying the history of the response $X(t;\theta)$, that multiply the random delta function inside averages. In the special case of initial value non-correlated with the excitation ($C_{X_0 \Xi(\bullet)}(t) = 0$), and zero-mean excitation ($m_{\Xi(\bullet)}(t) = 0$), Eq. (13) reduces to the SLE for RDEs under additive, coloured Gaussian excitation derived in various previous works [26,27,29,53], and [8], Eq. (3.27), where it is called the coloured noise master equation.

Before proceeding into deriving our new genFPK equation, we present four straightforward applications of SLE (13), establishing the consistency of our approach with existing methods. First, we rederive the classical FPK equation for the nonlinear RDE (1) under white noise excitation. Subsequently, moving on to coloured noise excitation cases, we derive an exact evolution equation for the response pdf of the linear RDE, as well as approximate genFPK equations for the nonlinear RDE (1) under small correlation time and Fox's approximations. For the last three cases, the equations derived herein are extended versions of existing genFPK equations, incorporating the effects of non-zero mean excitation and correlated initial value. In addition, the present methodology provides a simple and unifying way to obtain results that have been derived by various, usually more convoluted, ways in the literature; see e.g. [8,26,28,29].

(a) *Classical FPK equation for the nonlinear RDE under white noise excitation*

Consider SLE (13) for the nonlinear RDE, Eq. (1), under zero-mean white noise excitation:

$$m_{\Xi(\bullet)}(t) = 0, \qquad C^{\text{WN}}_{\Xi(\bullet)\Xi(\bullet)}(t,s) = 2D(t)\,\delta(t-s).$$

In this case, the upper time limit $t$ of the integral in the right-hand side of Eq. (13) coincides with the singular point of the delta function $\delta(t-s)$, making the value of this integral ambiguous. To resolve this ambiguity, we approximate the singular autocovariance function of the excitation, $C^{\text{WN}}_{\Xi(\bullet)\Xi(\bullet)}(t,s)$, by a weighted delta family,

$$C^{(\varepsilon)}_{\Xi(\cdot)\Xi(\cdot)}(t,s) = 2D(t)\,\delta_{\varepsilon}(t-s), \tag{14}$$

where $\delta_{\varepsilon}(t-s) = (1/\varepsilon)\,q((t-s)/\varepsilon)$, with $q(\cdot)$ being a non-negative smooth kernel function ([54], Ch. 20). In addition, in the present case, $q(\cdot)$ should be even, in order that $C^{(\varepsilon)}_{\Xi(\cdot)\Xi(\cdot)}(t,s)$ is a valid autocovariance. The last requirement implies that $\lim_{\varepsilon \downarrow 0} \int_{t_0}^{t} \delta_{\varepsilon}(t-s)\,ds = 1/2$, which, after the standard proof procedure for kernel functions ([55] Sec. 12.4, [54], *loc. cit.*), leads to identity

$$\lim_{\varepsilon \downarrow 0} \int_{t_0}^{t} \delta_{\varepsilon}(t-s)\,g(s)\,ds = \frac{1}{2}\,g(t), \tag{15}$$

for any continuous function $g(\cdot)$. Since $\exp(\mathcal{I}_{h'}[X(\cdot|_s^t;\theta)])$ is a continuous function with respect to its argument $s$, identity (15) can be employed, resulting in the following calculation of the integral in Eq. (13):

$$\int_{t_0}^{t} C^{WN}_{\Xi(\cdot)\Xi(\cdot)}(t,s)\exp(\mathcal{I}_{h'}[X(\cdot|_s^t;\theta)])\,ds \equiv \lim_{\varepsilon \downarrow 0}\int_{t_0}^{t} C^{(\varepsilon)}_{\Xi(\cdot)\Xi(\cdot)}(t,s)\exp(\mathcal{I}_{h'}[X(\cdot|_s^t;\theta)])\,ds =$$

$$= 2D(t)\lim_{\varepsilon \downarrow 0}\int_{t_0}^{t} \delta_{\varepsilon}(t-s)\exp(\mathcal{I}_{h'}[X(\cdot|_s^t;\theta)])\,ds = D(t)\exp(\mathcal{I}_{h'}[X(\cdot|_t^t;\theta)]) = D(t). \tag{16}$$

Substituting Eq. (16) into SLE (13), and assuming $C_{X_0\Xi(\cdot)}(t) = 0$, we obtain

$$\frac{\partial f_{X(t)}(x)}{\partial t} + \frac{\partial}{\partial x}\bigl(h(x)\,f_{X(t)}(x)\bigr) = \kappa^2 D(t)\frac{\partial^2 f_{X(t)}(x)}{\partial x^2}, \tag{17}$$

which is the classical FPK equation corresponding to the nonlinear RDE (1), excited by additive white noise.

**(b)** *Exact genFPK equation for the linear RDE under coloured noise excitation*

In case of a stable linear RDE under coloured noise excitation, $h(x) = \eta_1 x$, $\eta_1 < 0$, the term $\exp(\mathcal{I}_{h'}[X(\cdot|_s^t;\theta)])$ simplifies to the deterministic function $\exp(\eta_1(t-s))$. Thus, SLE (13) gives the following closed, *exact* genFPK equation:

$$\frac{\partial f_{X(t)}(x)}{\partial t} + \frac{\partial}{\partial x}\Bigl[\bigl(\eta_1 x + \kappa\,m_{\Xi(\cdot)}(t)\bigr) f_{X(t)}(x)\Bigr] = D^{\text{eff}}(t)\frac{\partial^2 f_{X(t)}(x)}{\partial x^2}, \tag{18}$$

where the term $D^{\text{eff}}(t)$, called the *effective noise intensity*, is given by

$$D^{\text{eff}}(t) = \kappa\,e^{\eta_1(t-t_0)}\,C_{X_0\Xi(\cdot)}(t) + \kappa^2 \int_{t_0}^{t} e^{\eta_1(t-s)}\,C_{\Xi(\cdot)\Xi(\cdot)}(t,s)\,ds. \tag{19}$$

Of course, in this case, the response pdf is already known to be a Gaussian one, which is easily calculated in terms of its mean value $m_{X(\cdot)}(t)$ and variance $\sigma^2_{X(\cdot)}(t)$. Nevertheless, Eq. (18) has a twofold value for the present work. First, it serves as a benchmark case for testing the accuracy and the efficiency of the numerical scheme developed for solving genFPK equations; second, due to its simple form, we are able to prove its unique solvability, a fact supporting the *conjecture* that the novel genFPK, Eq. (5), may be mathematically well-posed, as well. The fact that Eq. (18) admits the unique correct solution for any Gaussian excitation is proved rigorously in Appendix B in ESM. The same result, albeit obtained by an alternative method, is also proved in our recent work [48], Sec. 7. Note that, the probabilistic solution to the general linear problem can be effectively studied using characteristic functional techniques; see e.g. [56,57].

(c) *Small correlation time genFPK equation*

Moving on to nonlinear RDEs under additive coloured noise excitation, the non-local term $\exp(\mathcal{I}_{h'}[X(\bullet|_s^t;\theta)])$, considered as a function of $s$, can be approximated by a first order Taylor series around current time $t$:

$$\exp(\mathcal{I}_{h'}[X(\bullet|_s^t;\theta)]) \cong 1 + h'(X(t;\theta))(t-s). \tag{20}$$

Approximation (20) is valid for small correlations times of excitation and small cross-correlation times between excitation and the initial value, in which case the main effects of $C_{\Xi(\cdot)\Xi(\cdot)}(t,s)$, $C_{X_0\Xi(\cdot)}(u)$ are concentrated around current time $t$, and initial time $t_0$, respectively. Substitution of Eq. (20) into SLE (13), results in

$$\frac{\partial f_{X(t)}(x)}{\partial t} + \frac{\partial}{\partial x}\left[\left(h(x) + \kappa m_{\Xi(\cdot)}(t)\right) f_{X(t)}(x)\right] = \frac{\partial^2}{\partial x^2}\left[\left(D_0(t) + D_1(t) h'(x)\right) f_{X(t)}(x)\right], \tag{21}$$

in which the coefficients $D_0(t)$, $D_1(t)$ are given by the relation

$$D_n(t) = \kappa\, C_{X_0\Xi(\cdot)}(t)(t-t_0)^n + \kappa^2 \int_{t_0}^{t} C_{\Xi(\cdot)\Xi(\cdot)}(t,s)(t-s)^n\, ds, \text{ for } n = 0, 1. \tag{22}$$

Eq. (21) extends the *time-dependent small correlation time* (SCT) *genFPK equation* of Sancho, Hänggi and other authors [26], [8] Sec. V.A, to the case of correlated initial value and nonzero mean Gaussian excitation having general autocorrelation function (not only Ornstein-Uhlenbeck excitation). Note that, for the linear case, $h'(x) = \eta_1$, diffusion coefficient of SCT genFPK Eq. (21) equals to an approximation of $D^{\text{eff}}(t)$ of Eq. (18), obtained under a first order Taylor expansion of term $e^{\eta_1(t-s)}$ around $t$ with respect to $s$. Thus, SCT genFPK Eq. (21) fails to yield the exact genFPK Eq. (18) for the linear case. Furthermore, because of the approximation Eq. (20), diffusion coefficient of Eq. (21) may become negative. This unphysical feature, which limits the validity of Eq. (21) to the small correlation time regime, is discussed in detail in [26].

For the special case of deterministic initial value and zero-mean Ornstein-Uhlenbeck (OU) excitation ($m_{\Xi(\cdot)}(t) = 0$, $C_{\Xi(\cdot)\Xi(\cdot)}(t,s) = D e^{-|t-s|/\tau}/\tau$, $D > 0$), SCT genFPK Eq. (21) attains the stationary form,

$$\frac{\partial}{\partial x}\left(h(x) f_{X(t)}(x)\right) = D\kappa^2 \frac{\partial^2}{\partial x^2}\left[\left(1 + \tau h'(x)\right) f_{X(t)}(x)\right], \qquad (23)$$

called the *standard SCT genFPK equation* ([8], Eq. 5.6).

(d) *Fox's genFPK equation*

By approximating only the integral $\mathcal{I}_{h'}[X(\cdot|_s^t; \theta)]$ using a first order Taylor series around current time $t$, we obtain $\mathcal{I}_{h'}[X(\cdot|_s^t; \theta)] \cong h'(X(t;\theta))(t-s)$. Then, the nonlocal term becomes

$$\exp\left(\mathcal{I}_{h'}[X(\cdot|_s^t; \theta)]\right) \cong \exp\left[h'(X(t;\theta))(t-s)\right], \qquad (24)$$

which is the approximation proposed by Fox in [28]. Substituting Eq. (24) into SLE (13) results into the genFPK equation

$$\frac{\partial f_{X(t)}(x)}{\partial t} + \frac{\partial}{\partial x}\left[\left(h(x) + \kappa m_{\Xi(\cdot)}(t)\right) f_{X(t)}(x)\right] = \frac{\partial^2}{\partial x^2}\left(D(x,t) f_{X(t)}(x)\right), \qquad (25)$$

with diffusion coefficient $D(x,t)$

$$D(x,t) = \kappa C_{X_0 \Xi(\cdot)}(t) \exp\left(h'(x)(t-t_0)\right) + \kappa^2 \int_{t_0}^{t} C_{\Xi(\cdot)\Xi(\cdot)}(t,s) \exp\left(h'(x)(t-s)\right) ds. \quad (26)$$

Note that, contrary to the SCT Eq. (21), genFPK Eq. (25) is exact in the linear case. What is more, the diffusion coefficient $D(x,t)$ is now always positive, as required for physical (interpretation) and mathematical (well-posedness) reasons.

GenFPK Eq. (25) constitutes a time-domain extension of (stationary) Fox's genFPK equation, to non-zero mean Gaussian excitation with correlated initial value. By considering deterministic initial value, OU excitation, and $\tau h'(x) < 1$ (SCT condition), genFPK Eq. (25) gives rise to Fox's stationary genFPK equation [28]:

$$\frac{\partial}{\partial x}\left(h(x) f_{X(t)}(x)\right) = D\kappa^2 \frac{\partial^2}{\partial x^2}\left[\frac{1}{1 - \tau h'(x)} f_{X(t)}(x)\right]. \qquad (27)$$

Thus, by using the present approach, Fox's genFPK equation is rigorously rederived, without the ambiguities and controversies of the original path-integral approach employed by Fox in [28]. For a discussion on Fox's derivation and its relation to other derivations of genFPK equations, see e.g. [29]. Note also that, by formally expanding the term $1/(1 - \tau h'(x))$ of Eq. (27) in terms of $\tau$ and keeping only up to the linear term, $1/(1 - \tau h'(x)) \cong 1 + \tau h'(x)$, the stationary SCT genFPK Eq. (23) is retrieved.

## 3. Novel genFPK equations under Volterra adjustable decoupling approximation

In this section, we present the main novel result of the present work, called the Volterra adjustable decoupling approximation (VADA). In this approach, the nonlocal term of SLE (13) is approximated in two steps; first, the nonlocal term, $\exp(\mathcal{I}_{h'}[X(\bullet|_s^t;\theta)])$, is factorized in terms containing the various types of nonlinearities of $h(x)$ and, second, each of the said terms is approximated by an appropriate stochastic Volterra-Taylor functional series expansion around a certain instantaneous response moment.

Without loss of generality, we represent the nonlinear restoring function $h(x)$ as

$$h(x) = \eta_1 x + \sum_{k=2}^{N} \eta_k g_k(x) \equiv \sum_{k=1}^{N} \eta_k g_k(x)(^4), \quad \text{with } g_1(x) = x, \tag{28}$$

where $g_k(x)$, $k = 2, \ldots, N$, are given nonlinear functions, and $\eta_k$'s are constants. Under Eq. (28), the non-local term $\exp(\mathcal{I}_{h'}[X(\bullet|_s^t;\theta)])$, Eq. (12), can be split into

$$\exp\left(\mathcal{I}_{h'}[X(\bullet|_s^t;\theta)]\right) = \prod_{k=1}^{N} \exp\left(\eta_k \int_s^t g'_k(X(u;\theta))\,du\right). \tag{29}$$

Setting $Y_k(u;\theta) = g'_k(X(u;\theta))$ to simplify the writing, each term of the product in the right-hand side of Eq. (29), being a functional on $X(u;\theta)$ or $Y_k(u;\theta)$, is denoted as

$$\mathcal{G}_k\left[g'_k\left(X(\bullet|_s^t;\theta)\right)\right] \equiv \mathcal{G}_k[Y_k(\bullet|_s^t;\theta)] = \exp\left(\eta_k \int_s^t Y_k(u;\theta)\,du\right). \tag{30}$$

Note that $\mathcal{G}_1$, corresponding to the linear term of $h(x)$, is equal to $\exp(\eta_1(t-s))$, which does not need any further treatment. By the splitting of Eq. (29), the effect of each nonlinearity $g_k(x)$ on $\exp(\mathcal{I}_{h'}[X(\bullet|_s^t;\theta)])$ is encapsulated in only one $\mathcal{G}_k$. Thus, it is possible to approximate each $\mathcal{G}_k$ separately, being able to provide the appropriate treatment for each type of nonlinearity.

Each stochastic functional $\mathcal{G}_k[Y_k(\bullet|_s^t;\theta)]$ is approximated by a stochastic Volterra-Taylor functional series expansion *not around zero*, but around the deterministic mean value of its argument $R_{g'_k(\bullet)}(s) = \mathbb{E}^\theta\left[g'_k(X(s;\theta))\right] = \mathbb{E}^\theta\left[Y_k(s;\theta)\right]$:

$$\mathcal{G}_k[Y_k(\bullet|_s^t;\theta)] = \sum_{m=0}^{\infty} \frac{1}{m!} \int_s^t \overset{(m)}{\cdots} \int_s^t \hat{Y}_k(s_1;\theta) \cdots \hat{Y}_k(s_m;\theta) \frac{\delta^m \mathcal{G}_k[R_{g'_k(\bullet)}(\bullet|_s^t)]}{\delta Y_k(s_1;\theta) \cdots \delta Y_k(s_m;\theta)} ds_1 \cdots ds_m, \tag{31}$$

---

($^4$) The simplest choice of the nonlinear functions $g_k(x)$, which is considered in detail subsequently, is $g_k(x) = x^k$. However, the method presented herein is more general, and can treat other types of nonlinearities, e.g., $g_4(x) = \sin(ax)$ and/or $g_5(x) = \exp(-a^2 x^2)$.

where $\hat{Y}_k(s;\theta) = Y_k(s;\theta) - R_{g'_k(\bullet)}(s)$ denotes the stochastic fluctuation of $Y_k(s;\theta)$ around its mean value $R_{g'_k(\bullet)}(s)$.

**Remark 3.1:** The rationale behind the use of Volterra-Taylor series (31) for expanding stochastic functionals (30), is, in essence, to decompose the effects of nonlinearities into mean effects, and fluctuation effects. The first part (being nonlinear and nonlocal) is included in the final equation *without any approximation*, through the term $\mathcal{E}[R_{h'(\bullet)}(\bullet|_s^t)]$ in Eq. (40) below. Approximation is employed only on the second part, i.e. the effect of fluctuations around mean values. The importance of such a decomposition has been also recognized by van Kampen [9] and Terweil [58] for linear RDEs, where it was performed at the level of the RDE itself.

The Volterra derivatives appearing in the right-hand side of Eq. (31) can be analytically calculated, taking into account the form of the functional $\mathcal{G}_k$, Eq. (30). Using the rules of Volterra calculus [59,60], we find

$$\frac{\delta^m \mathcal{G}_k[R_{g'_k(\bullet)}(\bullet|_s^t)]}{\delta Y_k(s_1;\theta)\cdots \delta Y_k(s_m;\theta)} = \eta_k^m \, \mathcal{G}_k[R_{g'_k(\bullet)}(\bullet|_s^t)]. \tag{32}$$

Substituting the above result in Eq. (31), and reversing from the shorthand notation $Y_k(u;\theta)$ to its original equivalent $g'_k(X(u;\theta))$, we obtain the following expression of the functional $\mathcal{G}_k$ in terms of $X(u;\theta)$:

$$\mathcal{G}_k\left[g'_k\left(X(\bullet|_s^t;\theta)\right)\right] = \mathcal{G}_k[R_{g'_k(\bullet)}(\bullet|_s^t)] \times \\ \times \sum_{m=0}^\infty \frac{\eta_k^m}{m!} \int_s^t \cdots^{(m)} \int_s^t \left(g'_k(X_1(s_1;\theta)) - R_{g'_k(\bullet)}(s_1)\right)\cdots\left(g'_k(X(s_m;\theta)) - R_{g'_k(\bullet)}(s_m)\right) ds_1 \cdots ds_m. \tag{33}$$

For the case of polynomial nonlinearities, $g_k(x) = x^k$, Eq. (33) is specified into

$$\mathcal{G}_k\left[g'_k\left(X(\bullet|_s^t;\theta)\right)\right] = \mathcal{G}_k[R_{X(\bullet)}^{(k-1)}(\bullet|_s^t)] \times \\ \times \sum_{m=0}^\infty \frac{(k\eta_k)^m}{m!} \int_s^t \cdots^{(m)} \int_s^t \left(X^{k-1}(s_1;\theta) - R_{X(\bullet)}^{(k-1)}(s_1)\right)\cdots\left(X^{k-1}(s_m;\theta) - R_{X(\bullet)}^{(k-1)}(s_m)\right) ds_1 \cdots ds_m, \tag{34}$$

where $R_{X(\bullet)}^{(k)}(s)$ is now the $k$–th order moment of the response, $R_{X(\bullet)}^{(k)}(s) = \mathbb{E}^\theta\left[X^k(s;\theta)\right]$. To exploit Eq. (33), or Eq. (34), for the approximation of the nonlocal terms in the right-hand side of SLE (13), a current time approximation for the temporal integrals is needed. In particular, using the approximation

$$\int_s^t \left(g'_k(X(s_1;\theta)) - R_{g'_k(\bullet)}(s_1)\right) ds_1 \cong \left[g'_k(X(t;\theta)) - R_{g'_k(\bullet)}(t)\right](t-s), \tag{35}$$

and similarly for the other integrals([5]), we obtain

$$G_k\left[g'_k\left(X(\bullet|_s^t;\theta)\right)\right] \cong G_k[R_{g'_k(\bullet)}(\bullet|_s^t)] \cdot \sum_{m=0}^{\infty} \frac{1}{m!} \varphi_k^m(X(t;\theta))(t-s)^m, \qquad (36)$$

where

$$\varphi_k(x,t) = \varphi_k\left(x; R_{g'_k(\bullet)}(t)\right) = \eta_k\left(g'_k(x) - R_{g'_k(\bullet)}(t)\right). \quad ([6]) \qquad (37)$$

Note that, contrary to SCT and Fox's current time approximations, Eqs. (20), (24), current time approximation (35) is applied to integrals of the fluctuations $g'_k(X(t;\theta)) - R_{g'_k(\bullet)}(t)$ of random functions $g'_k(X(t;\theta))$ around their mean values, and not to the random functions per se. This fact makes Eq. (35) a more accurate current time approximation. By substituting Eq. (36) into Eq. (29), we obtain the approximation

$$\exp\left(I_{h'}[X(\bullet|_s^t;\theta)]\right) \cong \exp\left(\eta_1(t-s) + \sum_{k=2}^{N} \eta_k \int_s^t R_{g'_k(\bullet)}(u)\,du\right) \times \\ \times \prod_{k=2}^{N} \sum_{m=0}^{\infty} \frac{1}{m!} \varphi_k^m(X(t;\theta),t)(t-s)^m. \qquad (38)$$

The exponential term, in the right-hand side of Eq. (38), is identified (via Eq. (28)) as the expression $\exp\left(\int_s^t \mathbb{E}^\theta\left[h'(X(u;\theta))\right]du\right)$ and is denoted, for simplicity, by $\mathcal{E}[R_{h'(\bullet)}(\bullet|_s^t)]$. That is,

$$\mathcal{E}[R_{h'(\bullet)}(\bullet|_s^t)] = \exp\left(\int_s^t \mathbb{E}^\theta\left[h'(X(u;\theta))\right]du\right) = \exp\left(\eta_1(t-s) + \sum_{k=2}^{N}\eta_k \int_s^t R_{g'_k(\bullet)}(u)\,du\right), \qquad (39)$$

where $R_{h'(\bullet)}(u) = \mathbb{E}^\theta\left[h'(X(u;\theta))\right]$. Since the series in the right-hand side of Eq. (38) are absolutely convergent, use of the Cauchy product for the multiplication of more than two series leads to the following, more convenient, equivalent form:

$$\exp\left(I_{h'}[X(\bullet|_s^t;\theta)]\right) \cong \mathcal{E}[R_{h'(\bullet)}(\bullet|_s^t)] \cdot \left(1 + \sum_{m=1}^{\infty}(t-s)^m \sum_{|\alpha|=m} \frac{\varphi^\alpha(X(t;\theta),t)}{\alpha!}\right), \qquad (40)$$

where $\boldsymbol{\varphi}(x,t) = (\varphi_2(x,t),\ldots,\varphi_N(x,t))$ and $\boldsymbol{\alpha} = (\alpha_2,\ldots,\alpha_N)$ is a multi-index. Recall that, $|\boldsymbol{\alpha}| = a_2 + \cdots + a_N$, $\boldsymbol{\varphi}^\alpha(x,t) = \varphi_2^{a_2}(x,t)\cdots\varphi_N^{a_N}(x,t)$ and $\boldsymbol{\alpha}! = (a_2!)\cdots(a_N!)$.

The right-hand side of approximation (40) contains two factors; the first one, $\mathcal{E}[R_{h'(\bullet)}(\bullet|_s^t)]$, is a functional on the response moments over time history, while the second is a sum encapsulating

---

([5]) Note that the multiple integrals in Eqs. (33) and (34) can be written as products of single ones.

([6]) Both notations, $\varphi_k(x,t)$ and $\varphi_k(x; R_{g'_k(\bullet)}(t))$ will be used subsequently. The former in the derivation procedure and the latter in the final form of genFPK equations, to make clear the dependencies.

the effect of fluctuations of the nonlinear terms around the said moments. Thus, approximation (40) retains a certain amount of nonlocality, through the term $\mathcal{E}[R_{h'(\cdot)}(\bullet|_s^t)]$, a feature that constitutes the main difference between the present method and the existing ones (Fox's and SCT).

By truncating the series in Eq. (40) at a finite $m = M$, and substituting it in SLE (13), we obtain the following novel genFPK equation

$$\frac{\partial f_{X(t)}(x)}{\partial t} + \frac{\partial}{\partial x}\left[\left(h(x) + \kappa m_{\Xi(\cdot)}(t)\right) f_{X(t)}(x)\right] =$$

$$= \frac{\partial^2}{\partial x^2}\left[\left(D_0^{\text{eff}}\left[R_{h'(\cdot)}(\bullet|_{t_0}^t), t\right] + \sum_{m=1}^{M} D_m^{\text{eff}}\left[R_{h'(\cdot)}(\bullet|_{t_0}^t), t\right] \sum_{|\alpha|=m} \frac{\varphi^\alpha\left(x; \{R_{g'_k(\cdot)}(t)\}\right)}{\alpha!}\right) f_{X(t)}(x)\right], \quad (41)$$

in which coefficients $D_m^{\text{eff}}(t)$, called the *generalized effective noise intensities*, are given by

$$D_m^{\text{eff}}[R_{h'(\cdot)}(\bullet|_{t_0}^t), t] = \kappa\, \mathcal{E}[R_{h'(\cdot)}(\bullet|_{t_0}^t)]\, C_{X_0 \Xi(\cdot)}(t)\, (t-t_0)^m +$$

$$+ \kappa^2 \int_{t_0}^{t} \mathcal{E}[R_{h'(\cdot)}(\bullet|_s^t)]\, C_{\Xi(\cdot)\Xi(\cdot)}(t, s)\, (t-s)^m\, ds, \quad (42)$$

and

$$\boldsymbol{\varphi}\left(x; \{R_{g'_k(\cdot)}(t)\}\right) = \left(\varphi_2\left(x; R_{g'_2(\cdot)}(t)\right), \ldots, \varphi_N\left(x; R_{g'_N(\cdot)}(t)\right)\right). \quad (43)$$

**Remark 3.2:** Recalling Eq. (37), we observe that the $x$–dependence of the terms $\varphi_k\left(x; R_{g'_k(\cdot)}(t)\right)$ reflects the effects of the nonlinearities of the RDE, while the dependence on $R_{g'_k(\cdot)}(t)$ introduces a new type *local probabilistic nonlinearity*, since moment $R_{g'_k(\cdot)}(t)$ depends on the unknown response pdf $f_{X(t)}(x)$ at the current time $t$.

**Remark 3.3:** The dependence of $D_m^{\text{eff}}$ on $\mathcal{E}[R_{h'(\cdot)}(\bullet|_{t_0}^t)]$ gives rise to a *nonlocal probabilistic nonlinearity*, since it involves the time history of $f_{X(\cdot)}(x)$, from $t_0$ up to the current time $t$.

Taking into account Remarks 3.1 and 3.2, and defining $\sum_{|\alpha|=0} \varphi^\alpha / \alpha! = 1$, the $f_X$–dependent diffusion coefficient in VADA, Eq. (41), can be written in the following concise form

$$\mathcal{B}[f_X; x, t] = \sum_{m=0}^{M} D_m^{\text{eff}}\left[f_{X(\cdot)}(\bullet), t\right] \sum_{|\alpha|=m} \frac{\varphi^\alpha\left(x; f_{X(t)}(\bullet)\right)}{\alpha!}. \quad (44)$$

The probabilistic nonlocality and nonlinearity appearing through $\mathcal{B}[f_X; x, t]$ are inherited from the nonlocal term of the SLE, Eq. (4). The absence of such terms in the classical FPK equation is due to the fact that the nonlocal term of the SLE is fully localized in the case of white noise exci-tation; see Sec. 2(a). We thus conclude that VADA approach retains an amount of the original nonlocality and nonlinearity of the SLE, albeit of a tractable nature, through the history and the current-time values of certain response moments([7]).

Equations of FPK type whose coefficients depend on the unknown pdf arise in many fields of physics, e.g. in quantum systems [61,62], in problems with anomalous diffusion [63], in non-equilibrium thermodynamics [64] or in systems with long-range interactions [65]. A survey of such equations, commonly called nonlinear FPK equations, is presented in the book [49].

**Remark 3.4:** Another important observation, concerning our main new result, Eq. (41), is the following. If we keep only the zeroth-order term $D_0^{\text{eff}}[R_{h'(\cdot)}(\bullet|_{t_0}^{t}), t]$ in the diffusion coefficient, Eq. (41) reduces to the time-dependent genFPK equation derived by using Hänggi's ansatz, also called the *decoupling approximation*, see e.g. [8, Eq. 5.17,27,28]. Thus, our approach systematizes and generalizes Hänggi's decoupling approximation, using Volterra-Taylor series. This fact justifies the name *Volterra adjustable decoupling approximation* (VADA), coined by the present authors. Note that such a generalization of Hänggi's ansatz (without increasing the order of $x$–derivatives in the genFPK equation), was deemed not possible by Hänggi himself; see [8] p. 273.

## 4. First numerical results

In order to quantify the range of validity and assess the accuracy of the novel genFPK equations proposed in this work, a number of numerical simulations has been performed and briefly presented in this section. The numerical method used for solving the various versions of genFPK equations employs: **i)** a partition of unity finite element method (PUFEM) for the discretization of $f_{X(t)}(x)$ in the state space [66], **ii)** a Bubnov-Galerkin technique for deriving ODEs governing the evolution of $f_{X(t)}(x)$, and **iii)** a Crank-Nicolson scheme for solving the said ODEs in the time domain. A brief description of the numerical scheme is given in Appendix C in ESM. Similar numerical methods have been used for solving the standard FPK equation by Kumar and co-workers [67,68]. This approach, being free of the burden of inter-element continuity/smoothness problems (because of the use of partition of unity), seems promising for extension to higher dimensions, and a first step towards this goal has been taken by Sun and Kumar in [69] for the classical FPK equation.

A feature peculiar to our novel genFPK equations, calling for special numerical treatment, is their nonlinear/nonlocal character. This peculiarity is treated by a self-contained, iterative scheme as follows: the current-time values of the response moments, needed for the calculation of diffusion coefficient $\mathcal{B}[f_X; x, t]$, are estimated by extrapolation based on the two previous time steps, and then are improved by iterations at the current time. Usually one or two iterations suffice. The final values of these moments, for each time instant, are stored and used for the calculation of the nonlocal terms (time integrals) in $D_m^{\text{eff}}$, as required by Eqs. (39) and (42). A more detailed description of the numerical scheme is given in Appendix C in ESM.

---

([7]) Which are dictated by the structure of nonlinearity of the corresponding RDE.

The numerical results to be presented subsequently concern the response pdf of the nonlinear, bistable RDE

$$\dot{X}(t;\theta) = \eta_1 X(t;\theta) + \eta_3 X^3(t;\theta) + \kappa \Xi(t;\theta), \quad X(t_0;\theta) = X_0(\theta), \quad (45a,b)$$

with $\eta_1 > 0$, $\eta_3 < 0$. For this case, the excitation is considered a zero-mean OU process, and the initial value $X_0(\theta)$ is taken uncorrelated to the excitation. Then, by introducing the dimensionless variables [8] $\tilde{t} = \eta_1 t$, $\tilde{X} = X\sqrt{|\eta_3|/\eta_1}$, $\tilde{X}_0 = X_0\sqrt{|\eta_3|/\eta_1}$ and $\tilde{\Xi} = \kappa \Xi \sqrt{|\eta_3|/\eta_1^3}$, RDE (45) is expressed as

$$\dot{\tilde{X}}(\tilde{t};\theta) = \tilde{X}(\tilde{t};\theta) - \tilde{X}^3(\tilde{t};\theta) + \tilde{\Xi}(\tilde{t};\theta), \quad \tilde{X}(\tilde{t}_0;\theta) = \tilde{X}_0(\theta). \quad (46a,b)$$

Furthermore, the determination of autocorrelation function of the normalized $\tilde{\Xi}(\tilde{t};\theta)$ calls for the normalization of the correlation time $\tau_{cor}$ and intensity $D_{OU}$ of OU noise excitation. For this purpose, the relaxation time $\tau_{rel}$ of the homogeneous variant ($\Xi(t;\theta) = 0$) of Eq. (45) is chosen as reference time. Since homogeneous Eq. (45) is a Bernoulli equation, its relaxation time is found to be $\tau_{rel} = 1/(2\eta_1)$ (by studying the long-time behaviour of analytic solution). Thus, the dimensionless correlation time, called also *relative correlation time*, is defined as

$$\tilde{\tau} = \tau_{cor}/\tau_{rel} = 2\eta_1 \tau_{cor}. \quad (47)$$

On the basis of Eq. (47) and the definition of the normalized excitation $\tilde{\Xi}(\tilde{t};\theta)$, the autocorrelation function of the latter is expressed as

$$C_{\tilde{\Xi}(\cdot)\tilde{\Xi}(\cdot)}(\tilde{t},\tilde{s}) = \frac{\tilde{D}}{\tilde{\tau}}\exp\left(-\frac{2|\tilde{t}-\tilde{s}|}{\tilde{\tau}}\right), \quad (48)$$

with dimensionless intensity

$$\tilde{D} = 2\kappa^2 D_{OU}\frac{|\eta_3|}{\eta_1^2}. \quad (49)$$

As an example, we consider the dimensionless RDE (46) with $\tilde{D} = 1$, and $\tilde{\tau}$ taking values in the range $0.1-3.0$. That is, we study a *strongly nonlinear*, *bistable* case, under *strong random excitation* outside of the small-noise intensity regime [8,34], over a *wide range of relative correlation times*. Initial pdf is taken to be Gaussian with zero mean value and variance $\sigma = 0.6$. In Figure 1 we present the evolution of the response pdf for increasing values of the ratio $\tilde{\tau} = \tau_{cor}/\tau_{rel}$, as calculated by solving the genFPK equations based on:

    Small correlation time approximation, Eq. (21) (SCT),
    $0^{th}$-order VADA (Hänggi's ansatz), Eq. (41) with $M = 0$ (HAN),
    $2^{nd}$-order VADA, Eq. (41) with $M = 2$ (VADA-II) and
    $4^{th}$-order VADA, Eq. (41) with $M = 4$ (VADA-IV).

In the same figure, results obtained by Monte Carlo (MC) simulations are plotted, denoted by sim in the legends, for comparison purposes. The choice of even order VADA genFPK equations is made in order to ensure the global positivity of the corresponding diffusion coefficients.

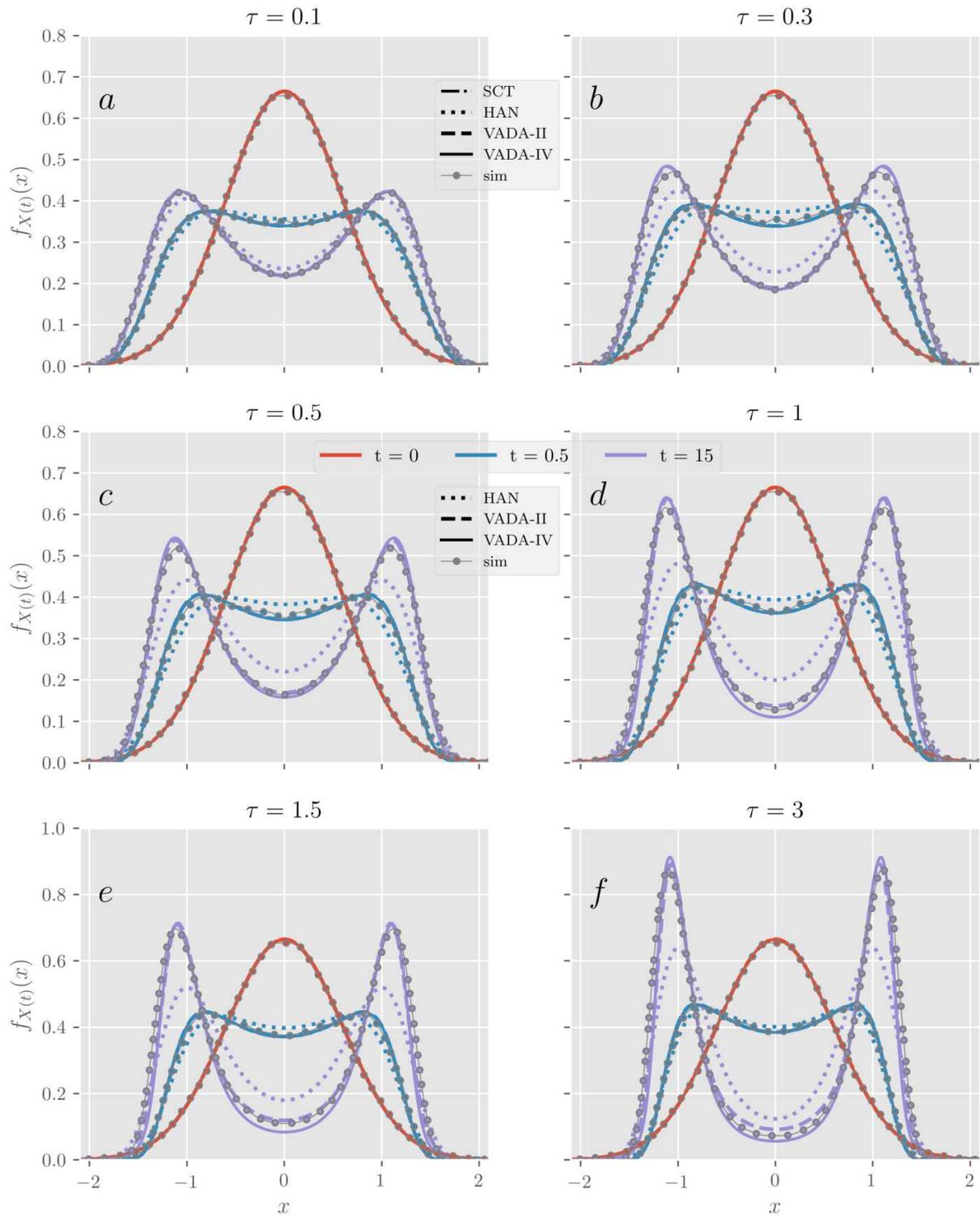

**Figure 1.** Evolution of response pdf for the RDE (46) excited by a zero-mean OU process with $D = 1$ and $\tau = 0.1, 0.3, 0.5, 1.0, 1.5, 3.0$. Initial pdf is zero-mean Gaussian with $\sigma = 0.6$. Results from various methods are presented along with MC simulations (coloured online).

Note that the final time instant in all plots is chosen to be in the long-time stationary regime, in order to check the validity of genFPK equations in both the transient and the stationary regimes.

As shown in the figure, for small values of $\tau$ (Fig. 1a, $\tau = \tau_{cor}/\tau_{rel} = 0.1$)([8]), all methods work well, with SCT and VADAs predicting a time evolution of response pdf in full agreement with MC simulations, both in the transient and the steady state regime. Only HAN slightly underestimates the peak values of the stationary pdf. As $\tau$ increases (Fig. 1b, $\tau = 0.3$) SCT is absent since the increase of $\tau$ renders its diffusion coefficient negative (and the SCT approximation invalid). HAN underestimates the pdf peak values more, while both VADA-II and IV are in almost full agreement with MC simulations. This picture is practically the same in Fig. 1c ($\tau = 0.5$), with HAN being even worse. For larger values of $\tau$ (Fig. 1d, $\tau = 1$), both VADAs are fairly accurate; however, they are a bit off at the peak values of the stationary pdf. For even larger values of $\tau$ (Fig. 1e, $\tau = 1.5$ and Fig. 1f, $\tau = 3$) HAN fails totally, while VADA-II and IV provide fairly accurate approximations, except for a minor failure at predicting the pdf peak values. Besides, the abscissae of the peak values predicted by VADAs are somewhat shifted closer to zero, in comparison with the ones of the MC results. In Appendix D in ESM, where a more detailed investigation of parameters is performed, it is shown that VADA-II and IV continue to give acceptable approximations of the pdf up to $D \approx 5$ and $\tau \approx 5$, with slightly but constantly increasing errors. As $D$ and $\tau$ increase further, the problem becomes more difficult to solve numerically; the numerical scheme, in its present form, exhibits instabilities or divergence towards the end of the transient state for various cases where $D\tau > 25$. The numerical solution of VADA-IV fails earlier than VADA-II (and HAN). This is reasonable, since higher-order polynomial terms are included in VADA-IV, amplifying any instabilities in the numerical scheme.

**Remark 4.1:** An interesting phenomenon observed in both Fig. 1 above and in Figs. 4a,b of Appendix D is that, for large values of $D$ and $\tau$, the response pdfs of RDE (46), obtained by MC simulations, exhibit their peak values at abscissae with absolute values larger than 1. This peak value drift, which has been documented before, see e.g. [8] p. 294, is predicted quite accurately by VADAs in the regime of $(D, \tau) = [0, 5.0] \times [0, 5.0]$, and -more generally- for $D\tau < 25$. VADA-IV is consistently more accurate than VADA-II. Note that this this peak value drift is not captured at all by HAN approach, which predicts pdfs with peaks fixed at $\pm 1$.

## 5. Conclusions and discussion

In the present work, we have developed a novel, efficient and accurate family of evolution equations governing the response pdf of a nonlinear scalar RDE, excited by coloured, Gaussian, additive noise. The excitation may have non-zero mean, and can be correlated with the initial state. The genFPK equations obtained are valid in both the transient and the stationary regimes, as well as for large correlation times and large noise intensities. What is more, their numerical solution, e.g. by using the PUFEM method, requires comparable computational effort with the corresponding classical FPK equation with the same state variables. Numerical results for a specific nonlinear bistable RDE confirm the validity and the accuracy of the proposed new genFPK equations.

---

([8]) From now on (in the text and in figures' captions) we omit the tilde from the non-dimensional quantities.

The derivation of the said new genFPK equations is made by elaborating the nonlocal term of the stochastic Liouville equations using: **i)** a new extended form of the Novikov-Furutsu theorem, **ii)** an approximation of the random nonlinear terms around their instantaneous moments, and **iii)** an expansion of the memory terms around current time. The obtained genFPK equations are nonlinear and nonlocal, yet easy to solve numerically. Also, they are able to rederive almost all existing genFPK equations (for additive coloured noise excitation) with appropriate simplifications of their terms. The approach presented herein has been called Volterra Adjustable Decoupling Approximation.

Finally, the present approach can be extended to systems of nonlinear RDEs under additive, coloured, Gaussian excitation. This direction is currently under consideration by the authors, and some first results were recently presented in [42]. More specifically, in [42], the multidimensional counterpart of SLE (13) is derived. In the multidimensional SLE, the nonlocal term -analogous to the nonlocal exponential term in SLE (13)- is identified as the transition matrix of the variational problem associated with the RDEs system. Subsequently, by appropriate approximations, we are able to easily re-derive the usual multidimensional genFPK equation, for additive noise excitation, found in the literature, see e.g. [70], as well as the multidimensional counterparts of Fox's genFPK equation, and Hänggi's ansatz, which to the best of our knowledge, have not been presented before. Multidimensional VADA genFPK equations have been also derived and will be submitted for publication soon. Apart from this, extension to multiplicative coloured noise excitation is possible, and it is under way, as well.

**Acknowledgements.** K.I. Mamis is supported by the ELKE-NTUA scholarship programme.

**Appendix A. Derivation of the stochastic Liouville equation via the delta projection method**

The random initial-value problem (RIVP) considered in the present work, Eq. (1a,b), is repeated here for easy reference:

$$\dot{X}(t;\theta) = h(X(t;\theta)) + \kappa \Xi(t;\theta), \tag{A1a}$$

$$X(t_0;\theta) = X_0(\theta). \tag{A1b}$$

The random data of the RIVP (A1) are the initial value $X_0(\theta)$ and the excitation $\Xi(t;\theta)$, fully described by the infinite-dimensional joint probability measure $\mathbf{P}_{X_0 \Xi(\bullet)}$, which is assumed to possess a well-defined probability density functional. The data measure $\mathbf{P}_{X_0 \Xi(\bullet)}$ is considered defined over the Borel $\sigma$–algebra of the product space $\mathbb{R} \times \mathscr{Z}$, where $\mathscr{Z}$ is with the space of continuous functions $C\big([t_0, t] \to \mathbb{R}\big)$. The marginal data measure $\mathbf{P}_{\Xi(\bullet)}$ is assumed to have continuous mean value and autocovariance operators (and thus continuous two-point correlation functions), reflecting the colored-noise character of the excitation $\Xi(t;\theta)$. The complete probabilistic solution of the RIVP (A1) is described by the joint response-excitation probability measure $\mathbf{P}_{X(\bullet) \Xi(\bullet)}$, which is assumed to exist and possess a joint probability density functional (and thus, finite-dimensional pdfs of all orders). It is assumed that the solution measure $\mathbf{P}_{X(\bullet) \Xi(\bullet)}$ is defined over the Borel $\sigma$–algebra of the product space $\mathscr{X} \times \mathscr{Z}$, where $\mathscr{Z}$ is the same as above, and $\mathscr{X}$ coincides with the space $C^1\big([t_0, t] \to \mathbb{R}\big)$, in which the response path functions belong. In addition, the solution measure $\mathbf{P}_{X(\bullet) \Xi(\bullet)}$ must obey the compatibility condition that its marginal measure $\mathbf{P}_{X(t_0) \Xi(\bullet)}$ is equal to the given data measure $\mathbf{P}_{X_0 \Xi(\bullet)}$. Having assumed that the underlying infinite-dimensional (measure) problem is well-posed, we focus on a systematic derivation of solvable approximate equations governing the evolution of the one-time response pdf $f_{X(t)}(x)$. Note that the response process $X(t;\theta)$ is non-Markovian, since the excitation $\Xi(t;\theta)$ is colored, and non-Gaussian when the system function $h(x)$ is nonlinear.

(a) *Delta projection method*

Representing the one-time pdf of a random function as the average of a random delta function,

$$f_{X(t)}(x) = \big\langle \delta(x - X(t;\theta)) \big\rangle, \tag{A2}$$

where $\langle \bullet \rangle$ is an appropriate ensemble average operator, widely-used practice in statistical mechanics (van Kampen, 2007, chap. XVI, sec.5), stochastic dynamics (Venturi *et al.*, 2012), and the theory of turbulence (Lundgren, 1967), where it is called the *pdf method*. Herein, the more suggestive term *delta projection method* is employed. In this subsection we (re)derive formula (A2) in a more generic way, obtaining also some generalizations of it which are useful in deriving the stochastic Liouville equation (SLE) corresponding to a nonlinear RDE.

To start with, we consider two integrable functions, $g_1(\bullet)$, $g_2(\bullet)$, and formulate the average of the product $g_1(X(\tau;\theta)) g_2(\Xi(s;\theta))$, for some (fixed) $s, \tau \in [t_0, t]$:

$$\mathbb{E}^\theta_{\mathbf{P}_{X(\bullet)\Xi(\bullet)}} \left[ g_1(X(\tau;\theta)) g_2(\Xi(s;\theta)) \right] = \\ = \int_{\mathscr{X} \times \mathscr{Z}} g_1(\chi(\tau)) g_2(\xi(s)) \mathbf{P}_{X(\bullet)\Xi(\bullet)}(d\chi(\bullet) \times d\xi(\bullet)). \quad (A3)$$

From now on the symbol $\mathbb{E}^\theta_{\mathbf{P}_{X(\bullet)\Xi(\bullet)}}[\bullet]$, of the ensemble average operator with respect to the joint response-excitation probability measure, is simplified to $\mathbb{E}^\theta[\bullet]$. Since the integrand in the right-hand side of Eq. (A3) depends only on the specific values $\chi(\tau)$ and $\xi(s)$ of the path functions $\chi(\bullet)$ and $\xi(\bullet)$, the infinite-dimensional integral is reduced to a two-dimensional one, with respect to marginal, two-point measure $\mathbf{P}_{X(\tau)\Xi(s)}$:

$$\mathbb{E}^\theta \left[ g_1(X(\tau;\theta)) g_2(\Xi(s;\theta)) \right] = \int_{\mathbb{R}^2} g_1(w) g_2(z) \mathbf{P}_{X(\tau)\Xi(s)}(dw \times dz),$$

which is also written, using the joint pdf $f_{X(\tau)\Xi(s)}(w, z)$, as

$$\mathbb{E}^\theta \left[ g_1(X(\tau;\theta)) g_2(\Xi(s;\theta)) \right] = \int_{\mathbb{R}^2} g_1(w) g_2(z) f_{X(\tau)\Xi(s)}(w, z) \, dw \, dz. \quad (A4)$$

We shall now apply Eq. (A4) to cases where one of the two functions $g_1(\bullet)$, $g_2(\bullet)$ is a generalized function, namely the delta function or some derivative of it. Justification of this extension can be made by invoking the theory of generalized stochastic processes; see e.g. (Gel'fand and Vilenkin, 1964, chap. III). Following the tradition in statistical physics, we shall proceed formally, without performing rigorous proofs in the context of the theory of generalized stochastic processes.

Setting $g_1(X(t;\theta)) = \delta(x - X(t;\theta))$ and $g_2(\Xi(s;\theta)) = 1$ in Eq. (A4), we obtain

$$\mathbb{E}^\theta \left[ \delta(x - X(t;\theta)) \right] = \int_{\mathbb{R}^2} \delta(x - w) f_{X(t)\Xi(s)}(w, z) \, dw \, dz = \\ = \int_{\mathbb{R}} \delta(x - w) f_{X(t)}(w) \, dw = f_{X(t)}(x), \quad (A5)$$

which is the same as Eq. (A2). Setting now

$$g_1(X(t;\theta)) = \frac{\partial \delta(x - X(t;\theta))}{\partial X(t;\theta)} q(X(t;\theta)) \quad \text{and} \quad g_2(\Xi(s;\theta)) = 1,$$

and assuming that the functions $q(x)$ and $f_{X(t)}(x)$ are continuously differentiable, Eq. (A4) provides us with the formula

$$\mathbb{E}^{\theta}\left[\frac{\partial \delta(x-X(t;\theta))}{\partial X(t;\theta)} q(X(t;\theta))\right] = \int_{\mathbb{R}^2} \frac{\partial \delta(x-w)}{\partial w} q(w) f_{X(t)\Xi(s)}(w,z) \, dw \, dz =$$

$$= \int_{\mathbb{R}} \frac{\partial \delta(x-w)}{\partial w} q(w) f_{X(t)}(w) \, dw =$$

$$= -\frac{\partial}{\partial x}\left(q(x) f_{X(t)}(x)\right). \tag{A6}$$

By replacing, in the above equation, the first derivative of the delta function by its $n^{th}$-derivative, and working similarly as above, we obtain the following useful generalization of Eq. (A6):

$$\mathbb{E}^{\theta}\left[\frac{\partial^n \delta(x-X(t;\theta))}{\partial X^n(t;\theta)} q(X(t;\theta))\right] = (-1)^n \frac{\partial^n}{\partial x^n}\left(q(x) f_{X(t)}(x)\right). \tag{A7}$$

For Eq. (A7) to be valid, the functions $q(x)$ and $f_{X(t)}(x)$ should possess $n^{th}$-order continuous derivatives. Finally, if we specify

$$g_1(X(t;\theta)) = \frac{\partial^n \delta(x-X(t;\theta))}{\partial^n X(t;\theta)} q(X(t;\theta)) \quad \text{and} \quad g_2(\Xi(t;\theta)) = \Xi(t;\theta),$$

and assume the same differentiability conditions on $q(x)$ and $f_{X(t)}(x)$ as above, Eq. (A4) leads to:

$$\mathbb{E}^{\theta}\left[\frac{\partial^n \delta(x-X(t;\theta))}{\partial X^n(t;\theta)} q(X(t;\theta)) \Xi(t;\theta)\right] = \int_{\mathbb{R}^2} \frac{\partial^n \delta(x-w)}{\partial w^n} q(w) z f_{X(t)\Xi(t)}(w,z) \, dw \, dz =$$

$$= (-1)^n \frac{\partial^n}{\partial x^n}\left(q(x) \int_{\mathbb{R}} z f_{X(t)\Xi(t)}(x,z) \, dz\right) =$$

$$= (-1)^n \frac{\partial^n}{\partial x^n}\left(q(x) \mathbb{E}^{\theta}\left[\delta(x-X(t;\theta)) \Xi(t;\theta)\right]\right). \tag{A8}$$

The last equality holds true since

$$\int_{\mathbb{R}} z f_{X(t)\Xi(t)}(x,z) \, dz = \int_{\mathbb{R}^2} z \delta(x-w) f_{X(t)\Xi(t)}(w,z) \, dw \, dz.$$

### (b) *Derivation of the stochastic Liouville equation*

We shall now apply the delta projection method to derive an equation governing the evolution of the response pdf $f_{X(t)}(x)$, when $X(t;\theta)$ satisfies RDE (A1a). By differentiating the first and the last members of Eq. (A5) with the respect to time, we find

$$\frac{\partial f_{X(t)}(x)}{\partial t} = \frac{\partial}{\partial t} \mathbb{E}^{\theta}\left[\delta(x-X(t;\theta))\right] = \mathbb{E}^{\theta}\left[\frac{\partial \delta(x-X(t;\theta))}{\partial X(t;\theta)} \dot{X}(t;\theta)\right]. \tag{A9}$$

The rightmost side of Eq. (A9) is derived by interchanging differentiation and expectation operators and using chain rule in differentiation. Now, in Eq. (A9), $\dot{X}(t;\theta)$ is substituted via RDE (A1a), leading to

$$\frac{\partial f_{X(t)}(x)}{\partial t} = \mathbb{E}^{\theta}\left[\frac{\partial \delta(x-X(t;\theta))}{\partial X(t;\theta)} h(X(t;\theta))\right] + \kappa \mathbb{E}^{\theta}\left[\frac{\partial \delta(x-X(t;\theta))}{\partial X(t;\theta)} \Xi(t;\theta)\right]. \qquad (A10)$$

Reformulation of the averages in the right-hand side of Eq. (A10), using Eqs. (A6) with $q(x) = h(x)$ for the first term, and Eq. (A8) with $n=1$ and $q(x)=1$ for the second term, results in

$$\frac{\partial f_{X(t)}(x)}{\partial t} + \frac{\partial}{\partial x}\left(h(x) f_{X(t)}(x)\right) = -\kappa \frac{\partial}{\partial x}\left(\mathbb{E}^{\theta}[\Xi(t;\theta) \delta(x-X(t;\theta))]\right), \qquad (A11)$$

which is the *stochastic Liouville equation* (SLE), Eq. (3) of the main paper. This equation has been derived by many authors, using various approaches; see e.g. (Hänggi, 1978; Sancho and San Miguel, 1980; Sancho *et al.*, 1982; Cetto et al., 1984; Fox, 1986; Venturi *et al.*, 2012). Also, the initial condition for SLE (A11) is easily determined, by use of Eq. (A5):

$$f_{X(t_0)}(x) = \mathbb{E}^{\theta}\left[\delta(x-X(t_0;\theta))\right] = \mathbb{E}^{\theta}\left[\delta(x-X_0(\theta))\right] = f_{X_0}. \qquad (A12)$$

In contrast with the classical Liouville equation (Schwabl, 2006, sec. 1.3.2), (van Kampen, 2007, sec. XVI.5), the SLE (A11) is not closed, due to the averaged term

$$\mathcal{N}_{\Xi X} = \mathbb{E}^{\theta}\left[\Xi(t;\theta) \delta(x-X(t;\theta))\right], \qquad (A13)$$

appearing in its right-hand side. This fact is better illustrated if the right-hand side of Eq. (A11) is rewritten as (see Eq. (8)):

$$\frac{\partial f_{X(t)}(x)}{\partial t} + \frac{\partial}{\partial x}\left(h(x) f_{X(t)}(x)\right) = -\kappa \int_{\mathbb{R}} z \frac{\partial f_{X(t)\Xi(t)}(x,z)}{\partial x} dz. \qquad (A14)$$

Eq. (A14) is clearly not closed, since, apart from $f_{X(t)}(x)$, it also contains the joint response-excitation pdf $f_{X(t)\Xi(t)}(x,z)$. Some ideas for obtaining approximate closures of the above equation have been presented in (Venturi *et al.*, 2012; Athanassoulis, Tsantili and Kapelonis, 2015). Nevertheless, our goal in this work, as mentioned in the introduction, is to provide a closure for the SLE in the form of Eq. (A11).

(c) *Comparison of SLE derivation presented herein to van Kampen's derivation*

Another technique for deriving SLE (A11) is by means of the so-called *van Kampen's lemma*; see (van Kampen 1975, p.269; van Kampen 1976, p.209; van Kampen 2007, p.411). The philosophy behind van Kampen's derivation of the SLE, based on the classical Liouville equation, is distinctively different from the one presented in the previous section, although it leads to the same result.

Van Kampen's derivation starts with a path-wise consideration of the given RDE. That is, the equation:

$$\dot{X}_\theta(t) = h(X_\theta(t)) + \kappa \Xi_\theta(t), \qquad X_\theta(t_0) = X_\theta^0, \tag{A15a,b}$$

is considered for every value of the stochastic argument $\theta$ separately. In this setting, (A15a,b) is a *deterministic* initial value problem. Then, the indicator function

$$p_\theta(x,t) = \delta(x - X_\theta(t)),$$

satisfies the *classical Liouville equation*

$$\frac{\partial p_\theta(x,t)}{\partial t} = -\frac{\partial}{\partial x}\left[\left(h(x) + \kappa \Xi_\theta(t)\right) p_\theta(x,t)\right], \tag{A16a}$$

$$p_\theta(x,t_0) = \delta(x - X_\theta^0); \tag{A16b}$$

(see (Klyatskin, 2005), Sec. 2.1). Van Kampen does not follow a formal derivation as Klyatskin's, using instead physical arguments after interpreting $p_\theta(x,t)$ as a "flow" and Eq. (A16a) as a conservation law. Although the argument of van Kampen is somewhat obscure, it can be justified by considering $p_\theta(x,t)$ as a counter of $\theta$–values (outcomes) satisfying the condition $X_\theta(t) = X(t;\theta) = x$ ("flow of outcomes").

By applying, on both Eqs. (A16a,b) the ensemble average operator $\langle \cdot \rangle$ over all $\theta$, we obtain:

$$\frac{\partial \langle p_\theta(x,t) \rangle}{\partial t} + \frac{\partial}{\partial x}\left(h(x)\langle p_\theta(x,t)\rangle\right) = -\kappa \frac{\partial}{\partial x}\langle \Xi(t;\theta) p_\theta(x,t)\rangle, \tag{A17a}$$

$$\langle p_\theta(x,t_0)\rangle = \langle \delta(x - X_0(\theta))\rangle. \tag{A17b}$$

The step that completes van Kampen's derivation of the SLE, see Eqs. (A11), (A12), is the identification

$$f_{X(t)}(x) = \langle p_\theta(x,t)\rangle. \tag{A18}$$

Eq. (A18) constitutes the van Kampen's lemma per se. In his works, (van Kampen, *loc. cit.*), van Kampen proves Eq. (A18) by employing a conservation argument. Note that, if we accept $\delta(x - X(t;\theta))$ as a generalized random function, then the justification of Eq. (A18) is straightforward, being a special case of Eq. (A4) of our approach:

$$\langle p_\theta(x,t)\rangle = \langle \delta(x - X(t;\theta))\rangle = \int_{\mathbb{R}} \delta(x-u) f_{X(t)}(u)\, du = f_{X(t)}(x). \tag{A19}$$

From the above discussion, a remarkable conceptual difference between van Kampen's Lemma and Eq. (A19) is revealed: in the former approach, the delta function is always treated as a deter-ministic function, with the probabilistic arguments being provided by the conservation law, while, in the latter, the delta function is interpreted, from the very beginning, as a stochastic function, on which the standard mean-value operator is applied.

In conclusion, although both approaches are interesting, revealing different conceptual settings, we come to believe that formula (A4), on which our approach is based, is an ample generalization of van Kampen's lemma, Eq. (A18), providing a rigorous and systematic way to treat various types of averages occurring in the derivation of the SLE.

**Appendix B. On the solvability of the genFPK equation corresponding to a linear RDE**

In this appendix we study the solvability (uniqueness and existence of solution) of the exact genFPK Eq. (18) of the main paper, governing the evolution of the one-time response pdf $f_{X(t)}(x)$ of a linear RDE under additive, Gaussian, coloured excitation. This issue seems to be trivial from the practical point of view, since in this case the response pdfs of all orders are known (Gaussian). Nevertheless, it is significant from the foundational point of view, since it ensures the well-posedness of a new equation, as well as its consistency with known results. In addition, after its theoretical justification, this equation provides us with a useful benchmark case for the assessment of the accuracy of the numerical methods for solving the various genFPK equations, as e.g. Eq. (21), (25) and (41) of the main paper, which are more complicated variants of the same type, that is, advection-diffusion equations.

(a) *Statement of the problem*

To facilitate the reader, we start by giving a complete yet concise statement of what has been already proved in Sec. 2 of the main paper, concerning the exact genFPK Eq. (18).

Under the conditions that: **i)** the excitation $\Xi(t;\theta)$ is a smoothly correlated (coloured) Gaussian process, with mean value $m_{\Xi(\cdot)}(t)$ and autocovariance $C_{\Xi(\cdot)\Xi(\cdot)}(t,s)$,

**ii)** the initial value $X_0(\theta)$ is a Gaussian random variable, with mean value $m_{X_0}$ and variance $\sigma^2_{X_0} \equiv C_{X_0 X_0}$, and

**iii)** $X_0(\theta)$ and $\Xi(t;\theta)$ are jointly Gaussian with cross-covariance $C_{X_0 \Xi(\cdot)}(t)$,

the one-time response pdf $f(x,t) = f_{X(t)}(x)$ of the linear random initial-value problem (RIVP):

$$\dot{X}(t;\theta) = \eta_1 X(t;\theta) + \kappa \Xi(t;\theta), \tag{B1a}$$

$$X(t_0;\theta) = X_0(\theta), \tag{B1b}$$

satisfies the following initial-value problem (IVP) (of a partial differential equation):

$$\frac{\partial f(x,t)}{\partial t} + \frac{\partial}{\partial x}\left[\left(\eta_1 x + \kappa m_{\Xi(\cdot)}(t)\right) f(x,t)\right] - D(t) \frac{\partial^2 f(x,t)}{\partial x^2} = 0, \tag{B2a}$$

$$t \geq t_0, \quad -\infty < x < +\infty,$$

$$f(x,t_0) = f_0(x) = given, \tag{B2b}$$

where the coefficient $D(t) = D^{\text{eff}}(t)$ (the effective noise intensity) is given by

$$D(t) = \kappa e^{\eta_1(t-t_0)} C_{X_0 \Xi(\cdot)}(t) + \kappa^2 \int_{t_0}^{t} e^{\eta_1(t-s)} C_{\Xi(\cdot)\Xi(\cdot)}(t,s) \, ds. \tag{B2c}$$

The IVP (B2) will be studied under the following plausible assumptions:

**A1)** The function $x \to f(x,t)$ belongs to $C^2(\mathbb{R})$ for each $t > t_0$,

The function $t \to f(x,t)$ belongs to $C^1((t_0, \infty))$ for each $x \in \mathbb{R}$,

The function $(x,t) \to f(x,t)$ is jointly continuous on $\mathbb{R} \times [t_0, \infty)$,

**A2)** The integrals $J(t) = \int_{-\infty}^{+\infty} f(x,t)\, dx$, $m_{X(\cdot)}(t) = \int_{-\infty}^{+\infty} x\, f(x,t)\, dx$,

$I(t) = \int_{-\infty}^{+\infty} f^2(x,t)\, dx$ and $P(t) = \int_{-\infty}^{+\infty} \left(\frac{\partial f(x,t)}{\partial x}\right)^2 dx$ <u>are finite</u>,

**A3)** $|x|\, f^2(x,t) \to 0$ as $|x| \to \infty$,

**A4)** $D(t) > 0$ for $t > t_0$.

A solution of the IVP (B2) satisfying the above assumptions is called a *classical solution*.

A first, straightforward consequence from Eq. (B2a) and the assumptions **A1)** and **A2)** is that the integral $J(t)$ is an invariant of the motion, which implies that $J(t) = J(t_0) = 1$, as it should be, since $f(x,t)$ is a probability density function. To prove this assertion, we integrate Eq. (B2a) over the real axis and observe that the quantities appearing as end terms in the integration by pats tend to zero as $|x| \to \infty$.

(b) *Uniqueness of solution*

**Theorem 1** [Uniqueness of the classical solution]. The IVP (B2) has at most one solution satisfying the assumptions **A1) – A4)**.

The main tool for proving Theorem 1 is the following

**Lemma 1.** Under the assumptions **A1) – A3)**, the integrals $I(t)$ and $P(t)$ of every classical solution of Eq. (B2a) satisfy the following identity, for any $t > t_0$:

$$\frac{1}{2}\frac{d}{dt} I(t) + \frac{1}{2} \eta_1 I(t) + D(t) P(t) = 0. \tag{B3}$$

*Proof* (of lemma): During the proof we simplify the notation, writing $m(t)$ instead of $m_{\Xi(\cdot)}(t)$. Multiplying both members of Eq. (B2a) by $f(x,t)$ and integrating over the whole $x$–axis, we obtain

$$\int_{-\infty}^{+\infty} \left( \frac{\partial f(x,t)}{\partial t} + \frac{\partial}{\partial x}\left[ (\eta_1 x + \kappa\, m(t))\, f(x,t) \right] - D(t) \frac{\partial^2 f(x,t)}{\partial x^2} \right) f(x,t)\, dx = 0.$$

(B4a)

All integrals appearing in the above equation are well defined on the basis of the stated assumptions. Eq. (B4a) is also written in the form

$$I_1(t) + I_2(t) + I_3(t) = 0, \tag{B4b}$$

where

$$I_1(t) = \int_{-\infty}^{+\infty} \frac{\partial f(x,t)}{\partial t} f(x,t)\, dx = \frac{1}{2} \frac{d}{dt} \int_{-\infty}^{+\infty} f^2(x,t)\, dx, \tag{B5a}$$

$$I_2(t) = \int_{-\infty}^{+\infty} \frac{\partial}{\partial x}\left[\left(\eta_1 x + \kappa\, m(t)\right) f(x,t)\right] f(x,t)\, dx =$$

$$= \eta_1 \int_{-\infty}^{+\infty} f^2(x,t)\, dx + \int_{-\infty}^{+\infty} \left(\eta_1 x + \kappa\, m(t)\right) f(x,t) \frac{\partial}{\partial x} f(x,t)\, dx =$$

$$= \eta_1 \int_{-\infty}^{+\infty} f^2(x,t)\, dx + \frac{1}{2} \int_{-\infty}^{+\infty} \left(\eta_1 x + \kappa\, m(t)\right) \frac{\partial f^2(x,t)}{\partial x}\, dx, \tag{B5b}$$

$$I_3(t) = -D(t) \int_{-\infty}^{+\infty} \frac{\partial^2 f(x,t)}{\partial x^2} f(x,t)\, dx. \tag{B5c}$$

Making an integration by parts in the last integral of the rightmost member of Eq. (B5b) and the integral in the right-hand side of Eq. (B5c), and observing that the end terms appearing after this integration by parts are zero because of the assumptions **A1) – A3)**, we find

$$\int_{-\infty}^{+\infty} \left(\eta_1 x + \kappa\, m(t)\right) \frac{\partial f^2(x,t)}{\partial x}\, dx = -\eta_1 \int_{-\infty}^{+\infty} f^2(x,t)\, dx, \tag{B6a}$$

$$\int_{-\infty}^{+\infty} \frac{\partial^2 f(x,t)}{\partial x^2} f(x,t)\, dx = -\int_{-\infty}^{+\infty} \left(\frac{\partial f(x,t)}{\partial x}\right)^2 dx. \tag{B6b}$$

Substituting the results (B6a,b) into Eqs. (B5b,c) we get

$$I_2(t) = \frac{1}{2} \eta_1 \int_{-\infty}^{+\infty} f^2(x,t)\, dx \tag{B7a}$$

and

$$I_3(t) = D(t) \int_{-\infty}^{+\infty} \left(\frac{\partial f(x,t)}{\partial x}\right)^2 dx. \tag{B7b}$$

Using Eqs. (B5a) and (B7a,b), Eq. (B4b) takes the form

$$\frac{1}{2} \frac{d}{dt} \int_{-\infty}^{+\infty} f^2(x,t)\, dx + \frac{1}{2} \eta_1 \int_{-\infty}^{+\infty} f^2(x,t)\, dx + D(t) \int_{-\infty}^{+\infty} \left(\frac{\partial f(x,t)}{\partial x}\right)^2 dx = 0,$$

which is identical with Eq. (B3). This completes the proof of the lemma.

We now proceed to the

*Proof of Theorem* 1. Assume that the IVP (B2) has two solutions, $f^{(1)}(x,t)$ and $f^{(2)}(x,t)$. Then, their difference $f^*(x,t) = f^{(1)}(x,t) - f^{(2)}(x,t)$ satisfies Eq. (B2a) and the homogeneous initial condition $f^*(x,t_0) = 0$. The integrals $I^*(t)$ and $P^*(t)$ (that is, the integrals $I(t)$ and $P(t)$ with $f$ substituted by $f^*$) will satisfy identity (B3), which is now written in the form

$$\frac{d}{dt} I^*(t) + \eta_1 I^*(t) = -2 D(t) P^*(t), \qquad t > t_0. \tag{B8}$$

Clearly,

$$P^*(t) \geq 0, \quad I^*(t) \geq 0 \quad \text{and} \quad I^*(t_0) = 0. \tag{B9a,b,c}$$

Considering Eq. (B8) as a differential equation with respect to $I^*(t)$, with initial condition (B9c), we obtain

$$I^*(t) = -2 \int_{t_0}^{t} D(\tau) P^*(\tau) \exp\left(-\eta_1 (t-\tau)\right) d\tau. \tag{B10}$$

Since $D(t) > 0$ and $P^*(t) \geq 0$, Eq. (B10) tells us that $I^*(t) \leq 0$. The latter inequality, in conjunction with inequality (B9b), implies that

$$I^*(t) \equiv \int_{-\infty}^{+\infty} f^{*2}(x,t) \, dx = 0.$$

Thus, $f^*(x,t) = 0$, which means that the two solutions, $f^{(1)}(x,t)$ and $f^{(2)}(x,t)$ are identical. This completes the proof of uniqueness theorem.

(c) *Existence of solution and consistency with known results*

We start by invoking the well-known result, that the response process of any linear system to an additive Gaussian excitation (either coloured or white) is also a Gaussian process. Thus, the response pdf $f_{X(t)}(x)$ of the linear RIVP (B1) is given by the formula

$$f_{X(t)}(x) = \frac{1}{\sqrt{2\pi \sigma_{X(\cdot)}^2(t)}} \exp\left[ -\frac{1}{2} \frac{(x - m_{X(\cdot)}(t))^2}{\sigma_{X(\cdot)}^2(t)} \right], \tag{B11}$$

where $m_{X(\cdot)}(t)$ is the mean value and $\sigma_{X(\cdot)}^2(t)$ is the variance of the response $X(t;\theta)$. The deterministic functions $m_{X(\cdot)}(t)$ and $\sigma_{X(\cdot)}^2(t)$ are defined as solutions of the corresponding moment equations which, in the present case, take the form (see e.g. (Sun, 2006) ch. 8, or (Athanassoulis, Tsantili and Kapelonis, 2015)):

$$\dot{m}_{X(\cdot)}(t) = \eta_1 m_{X(\cdot)}(t) + \kappa m_{\Xi(\cdot)}(t), \quad m_{X(\cdot)}(t_0) = m_{X_0}, \tag{B12a,b}$$

$$\frac{1}{2} \dot{\sigma}_{X(\cdot)}^2(t) = \eta_1 \sigma_{X(\cdot)}^2(t) + \kappa C_{X_0 \Xi(\cdot)}(t) e^{\eta_1(t-t_0)} +$$
$$+ \kappa^2 \int_{t_0}^{t} C_{\Xi(\cdot)\Xi(\cdot)}(u,t) e^{\eta_1(t-u)} du, \tag{B13a}$$

$$\sigma_{X(\cdot)}^2(t_0) = C_{X_0 X_0}. \tag{B13b}$$

Observe that, because of Eq. (B2c), Eq. (B13a) can be equivalently written as

$$\frac{1}{2}\dot{\sigma}^2_{X(\cdot)}(t) = \eta_1 \sigma^2_{X(\cdot)}(t) + D(t). \tag{B13a$'$}$$

Taking advantage of the above results, we shall now verify that the (unique) solution of the IVP (B2) is given by Eq. (B11). This establishes existence of solution, and consistency with existing results.

In order to verify that the function $f_{X(t)}(x)$, as defined by Eq. (B11), satisfies the IVP (B2), the temporal and spatial derivatives of it are needed. The calculation of these derivatives is straight-forward (details are omitted). The useful results are collected below:

$$\frac{\partial}{\partial t} f_{X(t)}(x) = \frac{f_{X(t)}(x)}{\left(\sigma^2_{X(\cdot)}(t)\right)^2} A\left(x\,;m_{X(\cdot)}(t),\sigma^2_{X(\cdot)}(t)\right), \tag{B14}$$

$$\frac{\partial}{\partial x} f_{X(t)}(x) = \frac{f_{X(t)}(x)}{\left(\sigma^2_{X(\cdot)}(t)\right)^2} B\left(x\,;m_{X(\cdot)}(t),\sigma^2_{X(\cdot)}(t)\right), \tag{B15}$$

$$\frac{\partial^2}{\partial x^2} f_{X(t)}(x) = \frac{f_{X(t)}(x)}{\left(\sigma^2_{X(\cdot)}(t)\right)^2} \Gamma\left(x\,;m_{X(\cdot)}(t),\sigma^2_{X(\cdot)}(t)\right), \tag{B16}$$

where

$$A = A\left(x\,;m_{X(\cdot)}(t),\sigma^2_{X(\cdot)}(t)\right) = \frac{1}{2}\dot{\sigma}^2_{X(\cdot)}(t)\, x^2 +$$
$$+ \left(\sigma^2_X(t)\,\dot{m}_{X(\cdot)}(t) - m_X(t)\,\dot{\sigma}^2_{X(\cdot)}(t)\right) x + \tag{B17}$$
$$+ \frac{1}{2} m^2_{X(\cdot)}(t)\,\dot{\sigma}^2_{X(\cdot)}(t) - m_{X(\cdot)}(t)\,\sigma^2_{X(\cdot)}(t)\,\dot{m}_{X(\cdot)}(t) - \frac{1}{2}\sigma^2_{X(\cdot)}(t)\,\dot{\sigma}^2_{X(\cdot)}(t).$$

$$B = B\left(x\,;m_{X(\cdot)}(t),\sigma^2_{X(\cdot)}(t)\right) = -\sigma^2_{X(\cdot)}(t)\, x + m_{X(\cdot)}(t)\,\sigma^2_{X(\cdot)}(t). \tag{B18}$$

$$\Gamma = \Gamma\left(x\,;m_{X(\cdot)}(t),\sigma^2_{X(\cdot)}(t)\right) = x^2 - 2\,m_{X(\cdot)}(t)\, x + m^2_{X(\cdot)}(t) - \sigma^2_{X(\cdot)}(t). \tag{B19}$$

Using Eqs. (B14) - (B16), the genFPK (advection-diffusion) Eq. (B2a) takes the form

$$A + \eta_1 \left(\sigma^2_{X(\cdot)}(t)\right)^2 + \left(\eta_1 x + \kappa\, m_{\Xi(\cdot)}(t)\right) B - D(t)\, \Gamma = 0\quad. \tag{B20}$$

**Lemma 2:** Eq. (B20) is satisfied for all $x \in \mathbb{R}$ and $t > t_0$ if and only if the moment equations (B12a) and (B13a$'$) are satisfied for all $t > t_0$.

*Proof*: Substituting the expressions (B17) – (B19), for the quantities $A, B, \Gamma$, into Eq (B20), and grouping together the coefficients of the same powers of $x$, results in

$$\left(\frac{1}{2}\dot{\sigma}^2_{X(\cdot)}(t) - \eta_1 \sigma^2_{X(\cdot)}(t) - D(t)\right) x^2 +$$

$$+ \left[\sigma^2_{X(\cdot)}(t) \left(\dot{m}_{X(\cdot)}(t) - \kappa m_{\Xi(\cdot)}(t)\right) - m_{X(\cdot)}(t) \left(\dot{\sigma}^2_{X(\cdot)}(t) - \eta_1 \sigma^2_{X(\cdot)}(t) - 2D(t)\right)\right] x +$$

$$+ \left(\frac{1}{2}\dot{\sigma}^2_{X(\cdot)}(t) - D(t)\right) m^2_{X(\cdot)}(t) + \left(\kappa m_{\Xi(\cdot)}(t) - \dot{m}_{X(\cdot)}(t)\right) m_{X(\cdot)}(t) \sigma^2_{X(\cdot)}(t) +$$

$$+ \left(-\frac{1}{2}\dot{\sigma}^2_{X(\cdot)}(t) + \eta_1 \sigma^2_{X(\cdot)}(t) + D(t)\right) \sigma^2_{X(\cdot)}(t) = 0. \qquad (B21)$$

This means that Eq. (B20) will be satisfied for all $x \in \mathbb{R}$ and $t > t_0$ if and only if the (three) coefficients of $x^2$, $x$ and $x^0$ (the $x$– independent term), in Eq. (B21), are zero for all $t > t_0$. We shall now show that the latter condition is equivalent with the satisfaction of the two moment equations (B12a) and (B13a′).

    Assume that Eqs. (B12a) and (B13a′) are satisfied. Then, the coefficient of $x^2$ in Eq. (B21) becomes zero, and the same happens with the last term of the coefficient of $x^0$. By substituting $\dot{\sigma}^2_{X(\cdot)}(t) = 2\eta_1 \sigma^2_{X(\cdot)}(t) + 2D(t)$, from Eq. (B13a′), in the coefficient of $x$, we obtain

Coeff. of $x = \sigma^2_{X(\cdot)}(t) \left(\dot{m}_{X(\cdot)}(t) - \eta_1 m_{X(\cdot)}(t) - \kappa m_{\Xi(\cdot)}(t)\right),$     (B22)

which vanishes because of Eq. (B12a). Finally, making the same substitution of $\dot{\sigma}^2_{X(\cdot)}(t)$ into the remaining part of the $x$– independent term, in Eq. (B21), we find

Coeff. of $x^0 = \left(\eta_1 m_{X(\cdot)}(t) + \kappa m_{\Xi(\cdot)}(t) - \dot{m}_{X(\cdot)}(t)\right) m_{X(\cdot)}(t) \sigma^2_{X(\cdot)}(t),$     (B23)

which also vanishes because of Eq. (B12a). This completes the proof that, if Eqs. (B12a) and (B13a′) are satisfied, then the coefficients of $x^2$, $x$ and $x^0$, in Eq. (B21), are zero and thus Eq. (B21) becomes an identity.

Assume now that the three coefficients of $x^2$, $x$ and $x^0$, in Eq. (B21), are zero. From the coefficient of $x^2$ we obtain Eq. (B13a′). Using the latter, we transform the coefficient of $x$ in the form of Eq. (B22), from which we conclude that Eq. (B12a) is also satisfied. Finally, the coefficients of $x^0$ takes again the form (B23), being zero without providing any new information. The proof of Lemma 2 is completed.

**Theorem 2** [Existence and consistency of solution]. The unique solution of the IVP (B2) is given by the Gaussian pdf, Eq. (B11).

*Proof*: According to Lemma 2, and the discussion preceding it, the function $f_{X(t)}(x)$, as defined by Eq. (B11), with mean value $m_{X(\cdot)}(t)$ and variance $\sigma^2_{X(\cdot)}(t)$ defined through the moment equations (B12a) and (B13a), satisfies Eq. (B2a). Further, the initial Gaussian pdf $f_0(x)$ is identical with $f_{X(t_0)}(x)$, since both have the same mean value $m_{X_0}$ and the same variance $\sigma^2_{X_0} \equiv C_{X_0 X_0}$; see Eqs. (B12b) and (B13b).

**Appendix C. Scheme of numerical solution. Implementation and validation**

The Partition of Unity Finite Element Method (PUFEM), introduced by Melenk and Babuška in 1996 (Melenk and Babuska, 1996), is a numerical method that generalizes the standard Finite Element Method (FEM) by utilizing the Partition of Unity (PU) concept. PUFEM has been already used for the numerical solution of the standard FPK equations, both in the stationary case (Kumar *et al.*, 2009) and in the transient case (Kumar, Chakravorty and Junkins, 2010). In this work, a PUFEM numerical scheme is adapted and implemented for the numerical solution of the genFPK equations derived in the main paper, including their non-linear, nonlocal variants.

(a) *Partition of Unity and PU approximation spaces*

**Theorem and Definition of Partition of Unity**: Let $\Omega \subset \mathbb{R}^n$ be an open domain, and $\{\Omega_k\}$, $k = 1(1)K \in \mathbb{N}$, be an open cover of $\Omega$ which satisfies *the pointwise overlap condition*:

$$(\exists M \in \mathbb{N})(\forall \boldsymbol{x} \in \Omega) : \operatorname{Card}\left(\{k_j\} \mid \boldsymbol{x} \in \Omega_{k_j}\right) \leq M. \quad (^9)$$

Then, for any given $s \in \mathbb{N}$, there exists a set of functions $\{\varphi_k(\boldsymbol{x})\}$, each one associated with an open domain $\Omega_k$, such that, for all $k = 1(1)K$:

1. $\varphi_k(\bullet) \in C^s(\mathbb{R}^n \to \mathbb{R})$,
2. $\operatorname{supp}(\varphi_k(\bullet)) \subseteq \overline{\Omega}_k$, $\quad(^{10})$
3. $(\forall \boldsymbol{x} \in \Omega)\ 0 \leq \varphi_k(\boldsymbol{x}) \leq 1$,

and

4. $(\forall \boldsymbol{x} \in \Omega)\ \sum_{k=1}^{K} \varphi_k(\boldsymbol{x}) = 1$.

We say that the set of functions $\{\varphi_k(\boldsymbol{x})\}$ forms a $C^s$–partition of unity ($C^s$–PU) associated with the cover $\{\Omega_k\}$. Functions $\varphi_k(\boldsymbol{x})$ are called partition of unity functions (PUF). ∎

For each subdomain $\Omega_k$ we consider an approximate function basis

$$\left\{b_\mu^k(\bullet) \in C^\ell(\Omega_k \to \mathbb{R}),\ \mu = 1, 2, \cdots, \mathrm{M}(k)\right\},$$

---

($^9$) Note that, for any $\boldsymbol{x} \in \Omega$, $\operatorname{Card}\left(\{k_j\} \mid \boldsymbol{x} \in \Omega_{k_j}\right) \geq 1$, since $\{\Omega_k\}$ is a cover of $\Omega$. In practice, the method is implemented so that $\operatorname{Card}\left(\{k_j\} \mid \boldsymbol{x} \in \Omega_{k_j}\right) \geq 2$ (except for a boundary "layer" near the physical boundary of the computational domain $\Omega$, where the said cardinality may be 1).

($^{10}$) From Conditions 1), 2) it easy to infer that $\operatorname{supp}(\varphi_k'(\xi)) = \operatorname{supp}(\varphi_k''(\xi)) = \operatorname{supp}(\partial^s \varphi_k(\xi)/\partial \xi^s) = \overline{\Omega}_k$.

where the necessary order of smoothness $\ell$ is dictated by the specific problem. The set $\{b_\mu^k(\bullet)\}$ is called the *local basis associated with* $\Omega_k$ (or the *local basis of* $\Omega_k$), and may contain different number of elements, $M(k)$, for different subdomains $\Omega_k$. In each $\Omega_k$, the local basis defines the *local approximation space*

$$V^{M_k}(\Omega_k) = \mathrm{span}\left(\{b_\mu^k(\boldsymbol{x})\}\right). \tag{C1}$$

In this work, local approximation spaces will be constructed by using (Legendre) orthogonal polynomials and, thus, they will be dense in $C(\Omega_k)$, $C^1(\Omega_k)$ and $H^1(\Omega_k)$.

The *approximate basis in the global domain* $\Omega$ is then constructed by means of the functions

$$u_\mu^k(\boldsymbol{x}) = \varphi_k(\boldsymbol{x})\, b_\mu^k(\boldsymbol{x}), \quad \boldsymbol{x} \in \Omega, \quad \begin{cases} k = 1, 2, \cdots, K, \\ \mu = 1, 2, \cdots, M(k). \end{cases} \tag{C2}$$

Following the FEM tradition, we shall call $u_\mu^k(\boldsymbol{x})$ *shape functions*. The span of shape functions defines the *global approximation space*, called also *partition of unity approximation space*, or *partition of unity space*, for short:

$$V^{\mathrm{PU}}(\Omega) = \mathrm{span}\left(\{u_\mu^k(\boldsymbol{x}) = \varphi_k(\boldsymbol{x})\, b_\mu^k(\boldsymbol{x}),\ k = 1, 2, \cdots, K,\ \mu = 1, 2, \cdots, M(k)\}\right). \tag{C3}$$

Given that the local approximation spaces $V^{M_k}(\Omega_k)$ have the necessary approximation properties, the global approximation space $V^{\mathrm{PU}}(\Omega)$ is also dense in $C(\Omega)$, $C^1(\Omega)$ and $H^1(\Omega)$ (see Theorem 2.1 in (Melenk and Babuska, 1996)), providing us with an efficient approximation of functions $F(\bullet)$ from the aforementioned spaces:

$$F(\boldsymbol{x}) \approx \hat{F}(\boldsymbol{x}) = \sum_{k=1}^{K} \sum_{\mu=1}^{M(k)} w_\mu^k\, u_\mu^k(\boldsymbol{x}), \quad \boldsymbol{x} \in \Omega, \tag{C4}$$

where $\hat{F}(\boldsymbol{x})$ denotes the PU approximation of $F(\boldsymbol{x})$ and $w_\mu^k$ are scalar weights.

**(b)** *Construction of cover,* PU *functions, and local basis functions for* 1D *domains*

We shall now present a specific implementation of the above concepts and constructions for one-dimensional (1D) domains. In our implementation $\mathrm{Card}\left(\{k_j\} \mid \boldsymbol{x} \in \Omega_{k_j}\right) = 2$ for almost all $\boldsymbol{x} \in \Omega$, with the exception of the "outer half" of the end subdomains ($\Omega_1$ and $\Omega_K$) where $\mathrm{Card}\left(\{k_j\} \mid \boldsymbol{x} \in \Omega_{k_j}\right) = 1$.

Let $\Omega = [\omega_{\text{g-min}}, \omega_{\text{g-max}}] \subset \mathbb{R}$ be a given domain (*the computational domain*) on the real axis, and $K$ be the number of subdomains of the cover $\{\Omega_k\}$ of $\Omega$. Then, each subdomain (interval) $\Omega_k = [\omega_{\min}^k, \omega_{\max}^k]$ is defined by means of the equations

$$\begin{aligned}\omega_{\min}^{k} &= \omega_{\text{g-min}} + (k-1)\cdot h, \\ \omega_{\max}^{k} &= \omega_{\text{g-min}} + (k+1)\cdot h = \omega_{\min}^{k} + 2h,\end{aligned} \quad (C5)$$

where $h = (\omega_{\text{g-max}} - \omega_{\text{g-min}})/(K+1)$. For this cover, the length of each subdomain is equal to $2h$, and the overlapping between adjacent subdomains is $h$. The layout of such an 1D cover is shown in Figure 1.

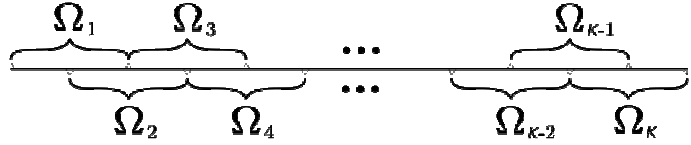

**Figure 1.** Layout of cover $\{\Omega_k\}$ of the computational domain $\Omega = \bigcup\limits_{k=1}^{K} \Omega_k$.

For the specification of PU functions and the local basis functions, it is more convenient to work in a *reference domain*, say $\Omega_{\text{ref}} = [-1, +1]$, and then transform the results into the physical domain. All calculations are performed in $\Omega_{\text{ref}}$ (the reference view) and converted to arbitrary $\Omega_k$ (the absolute view) by an affine map, defined by

$$\begin{aligned}\xi_k(\omega) &: [\omega_{\min}^{k}, \omega_{\max}^{k}] \to [-1, +1] \\ \xi_k(\omega) &= \frac{2\omega - \omega_{\min}^{k} - \omega_{\max}^{k}}{\omega_{\max}^{k} - \omega_{\min}^{k}}, \quad \omega \in [\omega_{\min}^{k}, \omega_{\max}^{k}]\end{aligned} \quad (C6)$$

with inverse map (from reference to absolute coordinates):

$$\begin{aligned}\omega_k(\xi) &: [-1, +1] \to [\omega_{\min}^{k}, \omega_{\max}^{k}] \\ \omega_k(\xi) &= \frac{(\omega_{\max}^{k} - \omega_{\min}^{k})\xi + \omega_{\max}^{k} + \omega_{\min}^{k}}{\omega_{\max}^{k} - \omega_{\min}^{k}}, \quad \xi \in [-1, 1]\end{aligned} \quad (C7)$$

We shall now construct the mother $C^s$–PU function in the reference domain $\Omega_{\text{ref}}$, which is denoted by $\tilde{\varphi}^{(s)}(\xi)$ [11]. This function is defined as a piecewise polynomial, by

$$\tilde{\varphi}^{(s)}(\xi) = \begin{cases} g_s(y_L(\xi)), & \text{for } \xi \in [-1, 0] \\ g_s(y_R(\xi)), & \text{for } \xi \in [0, 1] \\ 0, & \text{otherwise} \end{cases} \quad (C8)$$

where:

$$y_L(\xi) = 2\xi + 1, \qquad y_R(\xi) = -2\xi + 1, \quad (C9)$$

and $g_s(\bullet)$ is a polynomial satisfying the following conditions [12]

---

[11] Note that the superscript $(s)$ in $\tilde{\varphi}^{(s)}(\xi)$ indicates the order of smoothness of the PU functions (not to be confused with differentiation).

$$g_s(\xi) + g_s(-\xi) = 1, \quad \forall \xi \in \mathbb{R},$$

$$g_s(1) = 1,$$

$$\frac{d^\ell g_s(1)}{d\xi^\ell} = 0, \quad \text{for } \ell = 1(1)s.$$

It is only a matter of algebraic manipulations to show that such a polynomial exists, contains a constant term and only odd powers of the independent variable, and is of degree $2s-1$. Thus, its general form is

$$g_s(z) = a_0 + \sum_{i=1}^{i=s} a_i z^{2i-1}, \tag{C10}$$

with coefficients $a_i$ dependent on $s$ (values for cases $s = 1, 2, 3$ are shown in Table 1).

**Table 1.**: Polynomial coefficients for the PU function branch function (C10) with $s = 1, 2, 3$.

| Continuity | Coefficients |
|---|---|
| $C^1$ ($s = 1$) | $a_0 = \frac{1}{2}$, $a_1 = \frac{1}{2}$, $a_n = 0$ for $n \geq 2$ |
| $C^2$ ($s = 2$) | $a_0 = \frac{1}{2}$, $a_1 = \frac{3}{4}$, $a_2 = -\frac{1}{4}$, $a_n = 0$ for $n \geq 3$ |
| $C^3$ ($s = 3$) | $a_0 = \frac{1}{2}$, $a_1 = \frac{15}{16}$, $a_2 = -\frac{5}{8}$, $a_3 = \frac{3}{16}$, $a_n = 0$ for $n \geq 4$ |

For the local basis functions, the Legendre polynomials are used, usually defined by Rodrigues' formula, $P_n(\xi) = \frac{1}{2^n n!} \frac{d^n}{dx^n}(\xi^2 - 1)^n$. They are also expressed by the explicit formula

$$P_n(\xi) = 2^n \sum_{i=0}^{n} \xi^i \binom{n}{i} \binom{\frac{n+i-1}{2}}{n}, \qquad \xi \in \mathbb{R}. \tag{C11}$$

The local basis functions in $\Omega_{\text{ref}}$, denoted by $\tilde{b}_\mu(\xi)$, are defined by $\tilde{b}_\mu(\xi) = P_{\mu-1}(\xi)$. The first five of them are listed below

$$\tilde{b}_1(\xi) = P_0(\xi) = 1, \quad \tilde{b}_2(\xi) = P_1(\xi) = \xi, \quad \tilde{b}_3(\xi) = P_2(\xi) = \frac{1}{2}(3\xi^2 - 1),$$

$$\tilde{b}_4(\xi) = P_3(\xi) = \frac{1}{2}(5\xi^3 - 3\xi), \quad \tilde{b}_5(\xi) = P_4(\xi) = \frac{1}{8}(35\xi^4 - 30\xi^2 + 3).$$

Now, Eq. (C2) for the reference shape functions reads as follows

---

([12]) Sufficient conditions in order that Conditions 1) – 4) stated above are satisfied.

$$\tilde{u}_\mu(\xi) = \tilde{\varphi}^{(s)}(\xi) \cdot \tilde{b}_\mu(\xi). \tag{C12}$$

Figure 2 shows both the local basis functions and the corresponding shape functions (in the reference domain $\Omega_{\text{ref}}$) for smoothness of order $s = 2$. Note that $\tilde{u}_1(\xi) = \tilde{\varphi}^{(s)}(\xi)$.

Now, by using the affine transformation (C6), (C7), we are able to pass to the physical domain, obtaining the PU functions $\varphi_k^{(s)}(x)$, the local basis functions $b_\mu^k(x)$, and the shape functions $u_\mu^k(x)$, for any subdomain $\Omega_k$:

$$\varphi_k^{(s)}(x) = \tilde{\varphi}^{(s)}(\xi_k(x)), \qquad b_\mu^k(x) = \tilde{b}_\mu(\xi_k(x)) \tag{C13a,b}$$

$$u_\mu^k(x) = \varphi_k^{(s)}(x) b_\mu^k(x) = \tilde{u}_\mu(\xi_k(x)). \tag{C13c}$$

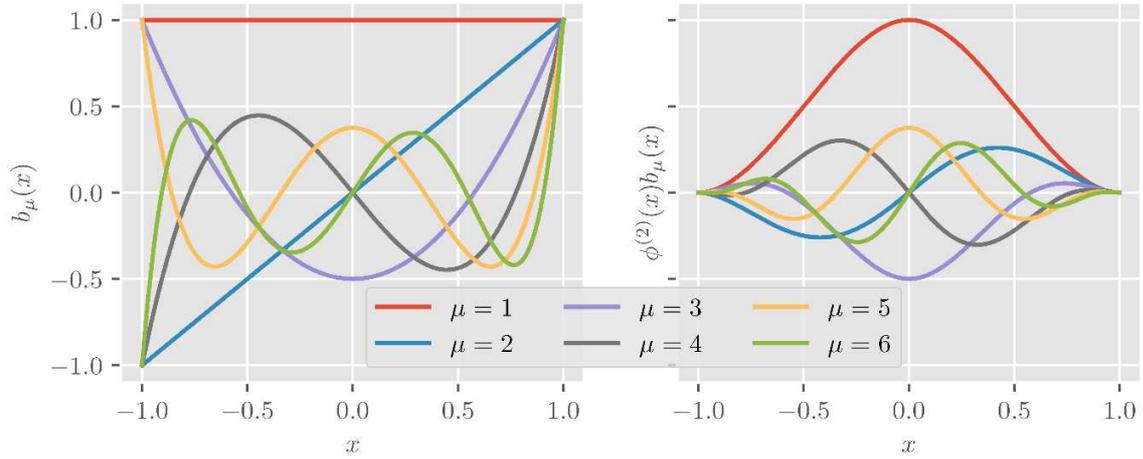

**Figure 2.** Legendre polynomial local basis functions $\tilde{b}_\mu(\xi)$ (left panel) and corresponding shape functions $\tilde{\varphi}^{(2)}(\xi) \cdot \tilde{b}_\mu(\xi)$ (right panel), for different values of $\mu$.

(c) *PUFEM discretization and numerical solution of pdf evolution equations*

Since the response pdf evolution equations, presented in Sec. 2 and 3 of the main paper, can attain a variety of forms, the following generic form is adapted for developing the numerical scheme

$$\frac{\partial f(x,t)}{\partial t} = -\frac{\partial}{\partial x}[q(x,t)f(x,t)] + \frac{\partial^2 \mathcal{B}[f;x,t]f(x,t)}{\partial x^2}, \tag{C14}$$

where now the unknown pdf is denoted by $f(x,t)$ instead of $f_{X(t)}(x)$. In Eq. (C14) $q(x,t)$ is the *drift coefficient*, and $\mathcal{B}[f;x,t]$ is the *diffusion coefficient*. Recall that the diffusion coefficient in VADA genFPK equations (41) depends on the unknown pdf $f(x,t)$, rendering Eq.

(C14) nonlinear and nonlocal.[13] Nevertheless, in this subsection, the simple notation $B(x,t)$ will be used for the diffusion coefficient, with notation $\mathcal{B}[f_X;x,t]$ being employed in the next subsection, where the treatment of nonlinearity/nonlocality will be explained.

Eq. (C14) is supplemented by the following conditions:

$$f(x,t_0) = f_0(x) = \text{known}, \quad \text{(initial condition)} \quad \text{(C15a)}$$
$$f(x,t) \geq 0, \quad \forall\, x \in \mathbb{R}, \quad \text{(positivity condition)} \quad \text{(C15b)}$$
$$f(x,t) \text{ decays sufficiently fast as } |x| \to \infty.\,[14] \quad \text{(decay condition)} \quad \text{(C15c)}$$

In addition, the unknown function $f(x,t)$ should satisfy the condition

$$\int_{-\infty}^{+\infty} f(x,t)\,dx = 1 \tag{C16}$$

since it is a probability density function. Note that this condition is automatically satisfied, since the initial value $f_0(x)$ is a pdf, and the integral (C16) is an invariant of Eq. (C14), as shown in Appendix B(a) of this paper.

A weak form of Eq. (C14) is obtained by projecting both members on the elements $v_j(x)$ of a test function space, $V^{\text{Test}}(\mathbb{R})$:

$$\frac{\partial}{\partial t}\int_{S(v_j)} f(x,t)\cdot v_j(x)\,dx = -\int_{S(v_j)} \frac{\partial}{\partial x}[q(x,t)f(x,t)]\cdot v_j(x)\,dx + \int_{S(v_j)} \frac{\partial^2 B(x,t)f(x,t)}{\partial x^2}\cdot v_j(x)\,dx, \tag{C17}$$

where $S(v_j) = \text{supp}(v_j(\cdot))$. Integration by parts on both terms in the right-hand-side of Eq. (C17) leads to the following equation, containing only first derivatives with respect to the unknown function $f(x,t)$:

$$\frac{\partial}{\partial t}\int_{S(v_j)} f(x,t)v_j(x)\,dx = \left(\frac{\partial B(x,t)f(x,t)}{\partial x}v_j(x) - q(x,t)f(x,t)v_j(x)\right)\bigg|_{\partial S(v_j)} + \int_{S(v_j)} \left(q(x,t) - \frac{\partial B(x,t)}{\partial x}\right)f(x,t)\frac{\partial v_j(x)}{\partial x}\,dx - \int_{S(v_j)} B(x,t)\frac{\partial f(x,t)}{\partial x}\frac{\partial v_j(x)}{\partial x}\,dx. \tag{C18}$$

To derive a discrete system of equations, the unknown solution $f(x,t)$ is approximated by using Eq. (C4):

$$f(x,t) \approx \hat{f}(x,t) = \sum_{k=1}^{K}\sum_{\mu}^{M(k)} w_\mu^k(t)u_\mu^k(x) = \sum_{m=1}^{M} w_m(t)u_m(x), \tag{C19}$$

---

[13] This is not the case for the small correlation time genFPK Eq. (21) of the main paper, whose diffusion coefficient does not depend on the unknown response pdf.
[14] The appropriate decay rate at infinity needs for ensuring the existence of the integrals over the $\mathbb{R}$, which appear in the weak form of Eq. (C10).

where $w_\mu^k(t) = w_m(t)$ are time-dependent weights, to be determined numerically, and the *global index* $m = m(k, \mu)$ is defined by:

$$m = m(k, \mu) = \sum_{i=1}^{k-1} M(i) + \mu \leq \sum_{i=1}^{K} M(i) \equiv M. \tag{C20}$$

(Recall that $M(i)$ denotes the number of basis functions in subdomain $\Omega_i$). Eq. (C20) can be inverted, to derive $k, \mu$ from $m$, by means of the formulae

$$k = k(m) = \max\left\{ j \in \mathbb{N} : \sum_{i=1}^{i=j} M(i) \leq m \right\},$$

$$\mu = \mu(m, k) = m - \sum_{i=1}^{i=k-1} M(i), \quad \left( \sum_{i=1}^{i=0} M(i) \equiv 0 \right).$$

The integer $M = \sum_{i=1}^{i=K} M(i)$ counts the total number of shape functions used in the discretization, and defines the *global Degree of Freedom* (DOF) of the scheme. To achieve better accuracy, the total DOF can be increased either by refining the cover $\{\Omega_k\}$ (which corresponds to $h$–FE refinement), or by increasing the number $M(k)$ of local basis functions in some (or all) subdomains $\Omega_k$ (which corresponds to $p$–FE refinement), or both ($hp$–FE refinement).

In all applications presented herein, the test space $V^{\text{Test}}(\mathbb{R})$ is chosen to be the same as the approximation space $V^{\text{PU}}(\Omega)$ (i.e. $u_m(x) = v_m(x), \forall m$), i.e. Bubnov-Galerkin approach (Hughes, 2000), leading to the following linear system:

$$\mathbf{C}\,\dot{\mathbf{w}}(t) = \mathbf{A}(t)\,\mathbf{w}(t), \tag{C21}$$

where:

$$\mathbf{C} \in \mathbb{R}^{M \times M} : c_{j,m} = \int_{S(u_m, u_j)} u_m(x)\,u_j(x)\,dx \tag{C22a}$$

$$\mathbf{A}(t): t \to \mathbb{R}^{M \times M} : a_{j,m} = \int_{S(u_m, u_j)} \left[ q(x,t) - \frac{\partial B(x,t)}{\partial x} \right] u_m(x) \frac{\partial u_j(x)}{\partial x}\,dx - \int_{S(u_m, u_j)} B(x,t) \frac{\partial u_m}{\partial x} \frac{\partial u_j(x)}{\partial x}\,dx, \tag{C22b}$$

where $S(u_m, u_j) = \text{supp}(u_m(\cdot)) \cap \text{supp}(u_j(\cdot))$. Note that the boundary terms of Eq. (C18) do not appear in Eq. (C22b), since they are all zero because of the properties of PU functions.

Approximating the time derivative in Eq. (C21) by means of a Crank-Nicolson scheme and grouping the terms with respect to their temporal argument, we obtain:

$$\left( \mathbf{C} - \frac{\Delta t}{2} \mathbf{A}(t + \Delta t) \right) \mathbf{w}(t + \Delta t) = \left( \frac{\Delta t}{2} \mathbf{A}(t) + \mathbf{C} \right) \mathbf{w}(t) \tag{C23}$$

where $\Delta t$ is the timestep. For each $t$, Eq. (C23) leads to an algebraic system of equations which can be solved to obtain $w(t+\Delta t)$ in terms of $w(t)$. Initialization of the numerical scheme requires knowledge of initial values of the weights $w(t_0) = w_0$, which are obtained by fitting the PU representation $\hat{f}(x, t_0)$ to the known initial data $f_0(x)$:

$$\mathbf{C}\,w_0 \;=\; \mathbf{f} \tag{C24}$$

where:

$$\mathbf{f} \in \mathbb{R}^M : f_j \;=\; \int_{S(u_j)} f_0(x)\, u_j(x)\, dx. \tag{C25}$$

Numerical solution of the linear systems (C23), (C24) is performed via the SciPy (Jones *et al.*, 2001) interface to LAPACK's (Anderson *et al.*, 1999) routine GESV. For solving the problem $AX = B$, GESV performs an LU-decomposition with partial pivoting and row interchanges to matrix $A$, namely $A = PLU$, where $P$ is a permutation matrix, $L$ is unit lower triangular, and $U$ is upper triangular; the factored form of matrix $A$ is used to solve the original system.

### (d) *Validation of the proposed scheme for the linear case*

Having described the numerical scheme for the solution of genFPK equations, we shall now present a first application to the simplest case that corresponds to the linear RDE under additive coloured noise excitation:

$$\dot{X}(t;\theta) \;=\; \eta_1 X(t;\theta) + \kappa \Xi(t;\theta), \qquad X(t_0;\theta) \;=\; X_0(\theta). \tag{C28a,b}$$

As it is derived in Section 2(b) of the main paper, the response pdf of RDE (C28) satisfies *exactly* the following genFPK equation (Eq. (18) of the main paper)

$$\frac{\partial f_{X(t)}(x)}{\partial t} + \frac{\partial}{\partial x}\left[\left(\eta_1 x + \kappa m_{\Xi(\cdot)}(t)\right) f_{X(t)}(x)\right] \;=\; D^{\text{eff}}(t)\, \frac{\partial^2 f_{X(t)}(x)}{\partial x^2}, \tag{C29}$$

with effective noise intensity

$$D^{\text{eff}}(t) \;=\; \kappa\, e^{\eta_1(t - t_0)}\, C_{X_0 \Xi(\cdot)}(t) + \kappa^2 \int_{t_0}^{t} e^{\eta_1(t - s)}\, C_{\Xi(\cdot)\Xi(\cdot)}(t, s)\, ds. \tag{C30}$$

As proved previously in Appendix B, exact genFPK (C29) has the Gaussian pdf

$$f_{X(t)}(x) \;=\; \frac{1}{\sqrt{2\pi \sigma^2_{X(\cdot)}(t)}}\, \exp\!\left[-\frac{1}{2}\, \frac{\left(x - m_{X(\cdot)}(t)\right)^2}{\sigma^2_{X(\cdot)}(t)}\right] \tag{C31}$$

as its unique solution, with $m_{X(\cdot)}(t)$, $\sigma^2_{X(\cdot)}(t)$ being the solutions to the linear ODEs (B12) and (B13) respectively, which can be easily expressed in closed form as

$$m_{X(\cdot)}(t) \;=\; m_{X_0}\, e^{\eta_1(t - t_0)} + \kappa \int_{t_0}^{t} m_{\Xi(\cdot)}(\tau)\, e^{\eta_1(t - \tau)}\, d\tau, \tag{C32}$$

and

$$\sigma_{X(\cdot)}^{2}(t) = C_{X_0 X_0} e^{2\eta_1(t-t_0)} + 2 \int_{t_0}^{t} D^{\text{eff}}(\tau) e^{2\eta_1(t-\tau)} d\tau. \tag{C33}$$

The numerical solution to genFPK Eq. (C29), obtained by using PUFEM with 200 degrees of freedom (50 PU functions, each with 4 basis functions), is compared to the analytic solution (C31) in Fig. 3 (left panels, a and c). Besides, in the right panels of Fig. 3, the approximate pdf $f_{X(t)}(x)$, obtained by direct Monte Carlo (MC) simulations of RDE (C28), is compared to the exact solution (C31). For the MC simulations, $5 \cdot 10^4$ realizations were solved and the pdf was approximated using a kernel density estimator (Scott, 2015). As can be seen from Fig.3, numerical results indicate that both the PUFEM numerical scheme and the MC simulations are able to accurately approximate the analytical solution. These results provide a first validation of the proposed new pdf equation, as well as the implementation of the PUFEM numerical scheme for its solution.

(e) *Treatment of the nonlinear, nonlocal terms*

As discussed above, a feature peculiar to VADA genFPK equations[15] is their nonlocality and nonlinearity, coming from the dependence of the diffusion coefficient on the unknown pdf through specific moments. The exact form of diffusion coefficient reads now as follows (see also Eqs. (41) - (44) of the main paper):

$$\mathcal{B}[f;x,t] = D_0^{\text{eff}}\left[ R_{h'(\cdot)}(\cdot|_{t_0}^{t}), t \right] + \sum_{m=1}^{M} D_m^{\text{eff}}\left[ R_{h'(\cdot)}(\cdot|_{t_0}^{t}), t \right] \sum_{|\boldsymbol{\alpha}|=m} \frac{\boldsymbol{\varphi}^{\boldsymbol{\alpha}}\left(x;\{R_{g_k'(\cdot)}(t)\}\right)}{\boldsymbol{\alpha}!}, \tag{C34}$$

where $R_{\zeta(\cdot)}(s) = \mathbb{E}^{\theta}\left[\zeta(X(s;\theta))\right]$ for any function $\zeta(x)$. As can be seen in Eq. (C34), $\mathcal{B}[f;x,t]$ depends on the current-time values of moments $R_{g_k'(\cdot)}(t)$, $k = 1, \ldots, M$, as well as on the whole time history[16] of moment $R_{h'(\cdot)}(\cdot)$, via coefficients $D_m^{\text{eff}}$:

$$D_m^{\text{eff}}[R_{h'(\cdot)}(\cdot|_{t_0}^{t}), t] = \kappa \exp\left(\int_{t_0}^{t} R_{h'(\cdot)}(u) \, du\right) C_{X_0 \Xi(\cdot)}(t) (t-t_0)^m +$$
$$+ \kappa^2 \int_{t_0}^{t} \exp\left(\int_{s}^{t} R_{h'(\cdot)}(u) \, du\right) C_{\Xi(\cdot)\Xi(\cdot)}(t,s) (t-s)^m \, ds. \tag{C35}$$

(Eq. (C35) is obtained from Eq. (42) in conjunction with Eq. (39) of the main paper). Thus, in each time step, we have to calculate the instantaneous values of the moments $R_{h'(\cdot)}(t)$, $R_{g_k'(\cdot)}(t)$, $k = 1, \ldots, M$ (generically denoted by $R(t)$), and the integral $\int_{s}^{t} R_{h'(\cdot)}(u) \, du$, which is done by means of a prediction-correction iterative scheme as follows:

---

[15] In the case of nonlinear RDEs.
[16] From initial time $t_0$ to current time $t$.

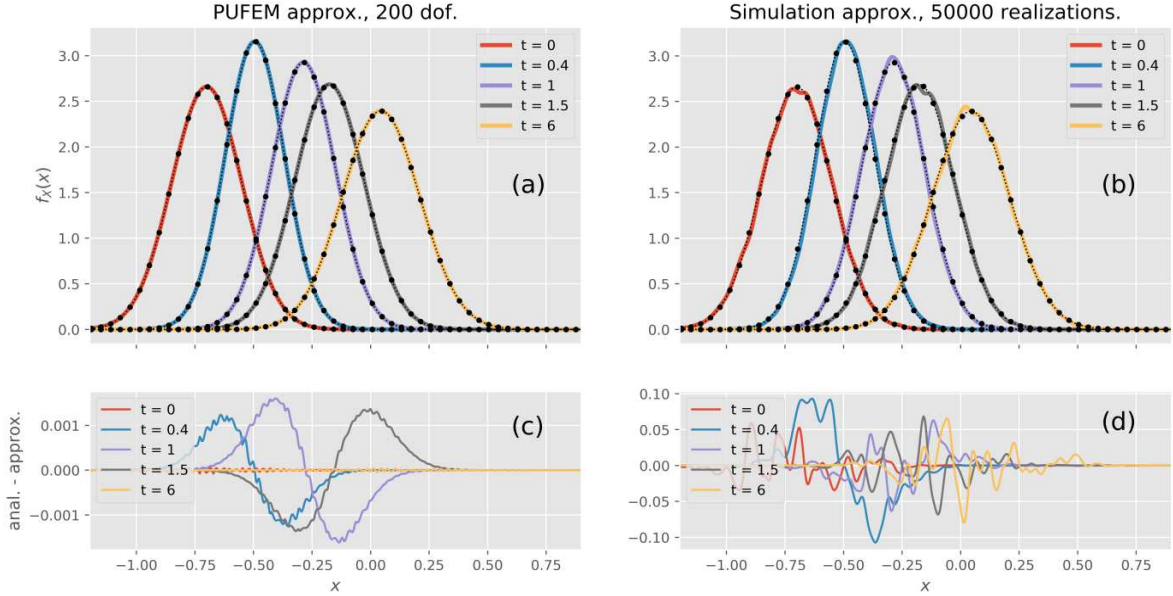

**Figure 3.** Evolution of response pdf for the linear RDE (C28): $\eta_1 = -0.8$, $\kappa = 0.2$, under non-zero mean Ornstein-Uhlenbeck excitation ($m_{\Xi(\cdot)} = 0.2$, $D = 1$, $\tau = 1$) and Gaussian initial data ($m_{X_0} = -0.7$, $\sigma_{X_0} = 0.15$). Analytical solutions (round markers in panels (a) and (b)) are compared to the PUFEM solution (panel (a)) and MC simulation approximation (panel (b)); errors of PUFEM and MC against the analytical solution are shown in panels (c) and (d), respectively.

**<u>Initial time step</u>** $i = 1$:

    **Calculate** $R(t_0)$ from the given initial pdf

    **Set** $R(t_1) = R(t_0)$

(\*)  **Set** $\int R(u)\,du = R(t_1)\,\Delta t$, where $\Delta t$ is the time step

    **Solve genFPK Eq. (C14)** using PUFEM to find pdf $f(x, t_1)$

    **Update** $R(t_1)$ to a new value $R_{\text{upd}}(t_1)$, by using $f(x, t_1)$

    **Check convergence condition** $\left| R(t_1) - R_{\text{upd}}(t_1) \right| \leq \varepsilon_{\text{tol}}$, for a given error $\varepsilon_{\text{tol}}$

    **If** convergence condition is **not satisfied**

        **Set** $R(t_1) = R_{\text{upd}}(t_1)$ and **return** to line (\*)

**<u>Time marching</u>** $i = 2, 3, \ldots, I$:

    **Calculate** $R(t_i)$ via linear extrapolation of $R(t_{i-1})$ and $R(t_{i-2})$

(&)  **Calculate** $\int R(u)\,du$ using the previous and the current values $R(t_k)$, $k = 0\,(1)\,i$

    **Solve genFPK Eq. (C14)** using PUFEM to find pdf $f(x, t_i)$

    **Update** $R(t_i)$ to a new value $R_{\text{upd}}(t_i)$, by using $f(x, t_i)$

**Check convergence condition** $\left| R(t_i) - R_{upd}(t_i) \right| \leq \varepsilon_{tol}$, for a given error $\varepsilon_{tol}$

**If** convergence condition is **not satisfied**

  **Set** $R(t_i) = R_{upd}(t_i)$ and **return** to line (&)

Note that, in every time step, the iterative scheme typically converges within only a few (1-2) iterations.

**Appendix D. Numerical Investigation of the range of validity of VADA genFPK equations**

As in Section 4 of the main paper, we will consider here the dimensionless cubic RDE([17])

$$\dot{X}(t;\theta) = X(t;\theta) - X^3(t;\theta) + \Xi(t;\theta), \qquad X(t_0;\theta) = X_0(\theta), \qquad \text{(D1a,b)}$$

under additive, zero-mean Ornstein-Uhlenbeck excitation $\Xi(t;\theta)$ with autocorrelation

$$C_{\Xi(\cdot)\Xi(\cdot)}(t,s) = \frac{D}{\tau}\exp\left(-\frac{2|t-s|}{\tau}\right), \qquad \text{(D2)}$$

where $D$ and $\tau$ are the dimensionless noise intensity and correlation time respectively. The procedure and justification for formulating dimensionless RDE (D1a,b) and dimensionless excitation autocorrelation (D2) can be found in Section 4 of the main paper.

By also considering that excitation $\Xi(t;\theta)$ is uncorrelated to initial value $X_0(\theta)$, the parameters governing the behaviour of RDE (D1a,b) solution are only the dimensionless $D$ and $\tau$ ([18]). In Figures 4a,b below, we show, for various values of $D$ and $\tau$, the stationary response pdfs of RDE (D1a,b) as calculated by solving the genFPK equations based on:

  $0^{th}$-order VADA (Hänggi's ansatz), Eq. (41) of the main paper with $M = 0$ (HAN),
  $2^{nd}$-order VADA, Eq. (41) of the main paper with $M = 2$ (VADA-II) and
  $4^{th}$-order VADA, Eq. (41) of the main paper with $M = 4$ (VADA-IV).

In the same Figures, the said stationary pdfs are compared with results obtained by Monte Carlo (MC) simulations, in order to determine the range of parameter values in which HAN, VADA-II and IV provide an accurate approximation.

For the MC simulations, the end of the transient state, $t_{st}$, is estimated by monitoring the stationarity (time invariance) of the first two moments of the response. From this time instant $t_{st}$ forward, we keep all simulation samples, with time step equal to the double of correlation time $\tau$ of excitation. This choice of time step ensures that we obtain uncorrelated samples. On the other hand, HAN, VADA-II and VADA-IV stationary pdfs are obtained by solving the respective time-dependent genFPK equations from the initial time up to their stationary regime. The end of the transient state of genFPK equations is taken equal to $t_{st}$ of MC simulations; then we check that their solutions remain unchanged for time equal to $2t_{st}$, to ensure that they are indeed stationary.

By observing Figs. 4a,b, we come up with the following three remarks:

**Remark D1: On the range of validity of Hänggi's ansatz.** For $D = 0.2$ and $1.0$ (Fig. 4a), HAN stationary pdfs are good approximations for $\tau = 0.1$ and $1.0$, while for larger noise intensity ($D = 2.0$ in Fig. 4b), the good performance of HAN method is restricted to the small correlation time regime ($\tau = 0.1$). For even larger noise intensities ($D = 5.0$ in Fig. 4b), HAN fails even for $\tau = 0.1$. In general, HAN method fails to capture the phenomenon of higher and narrower pdf peaks as the parameter $D\tau$ increases, resulting thus in pdfs that are more diffusive than the ones obtained by MC simulations. Roughly speaking, HAN becomes unreliable when $D\tau$ becomes larger than 1.

---

([17]) Expressed here without tilde over the dimensionless quantities.
([18]) A more detailed investigation would also include as parameters the (non-zero in general) mean values of excitation and initial value, as well as their cross-correlation.

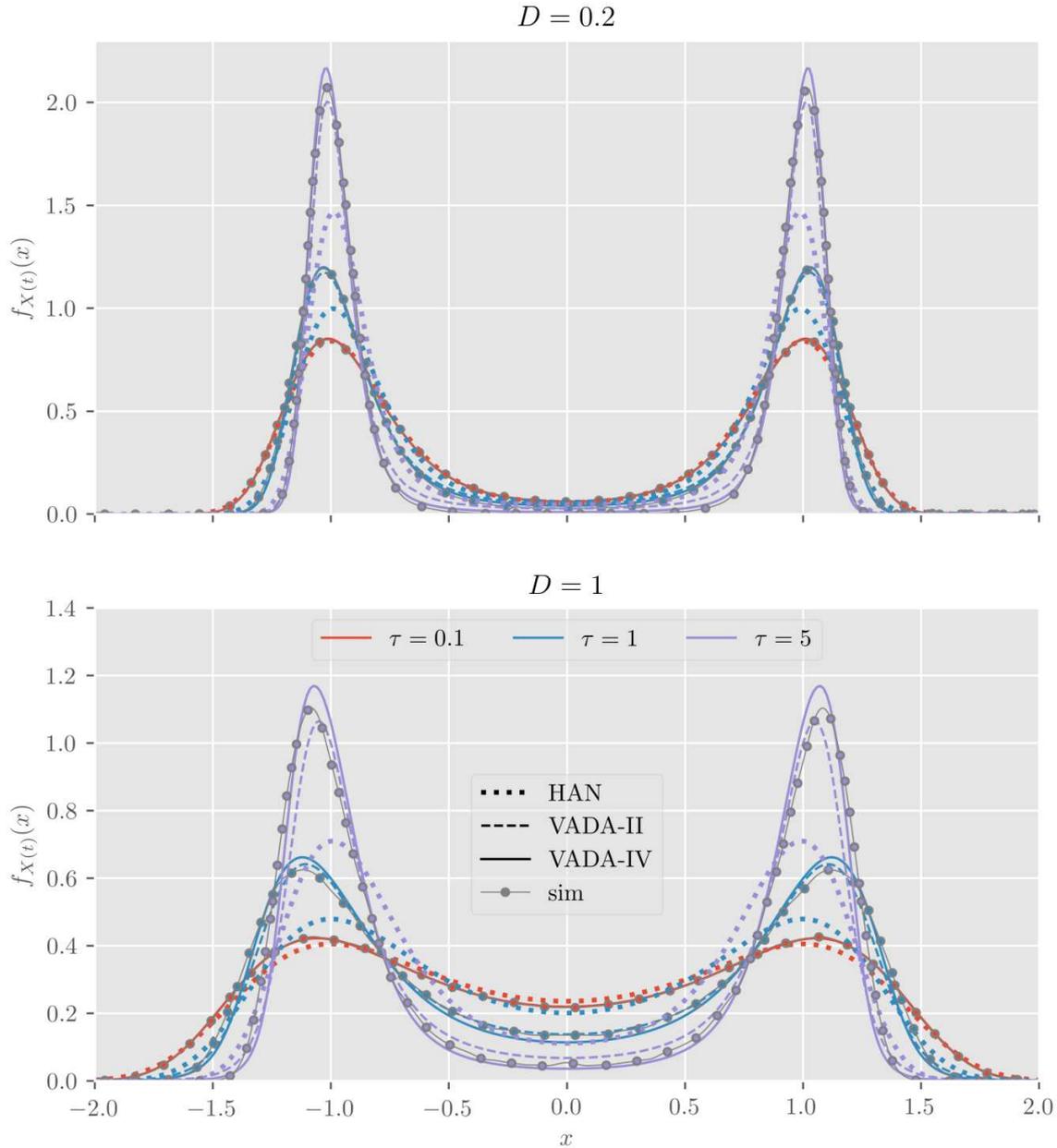

**Figure 4a.** Stationary response pdf for RDE (D1a,b) excited by a zero-mean OU process with $D = 0.2$ (upper panel) and $D = 1$ (lower panel) and for $\tau = 0.1, 1.0, 5.0$. Results from HAN, VADA-II and IV genFPK equations are presented along with MC simulations.

**Remark D2: On the range of validity of VADA-II and IV methods.** In Figs. 4a,b, we observe that the stationary pdfs obtained by solving VADA-II and IV genFPK equations are in good agreement with pdfs obtained by MC simulations, even in cases where both $D$ and $\tau$ attain large values. Note that, even for $D = 5.0$ and $\tau = 5.0$, VADA-II and IV continue to give acceptable approximations of the stationary pdf. A general trend is that, as $D$ and $\tau$ increase, VADA pdfs begin to fail at predicting the exact values and abscissae of the response pdf peaks. Roughly speaking, the range of validity of VADAs method seems to be limited by the condition $D\tau < 25$, providing a substantial improvement in comparison to HAN which is limited by the condition $D\tau < 1$.

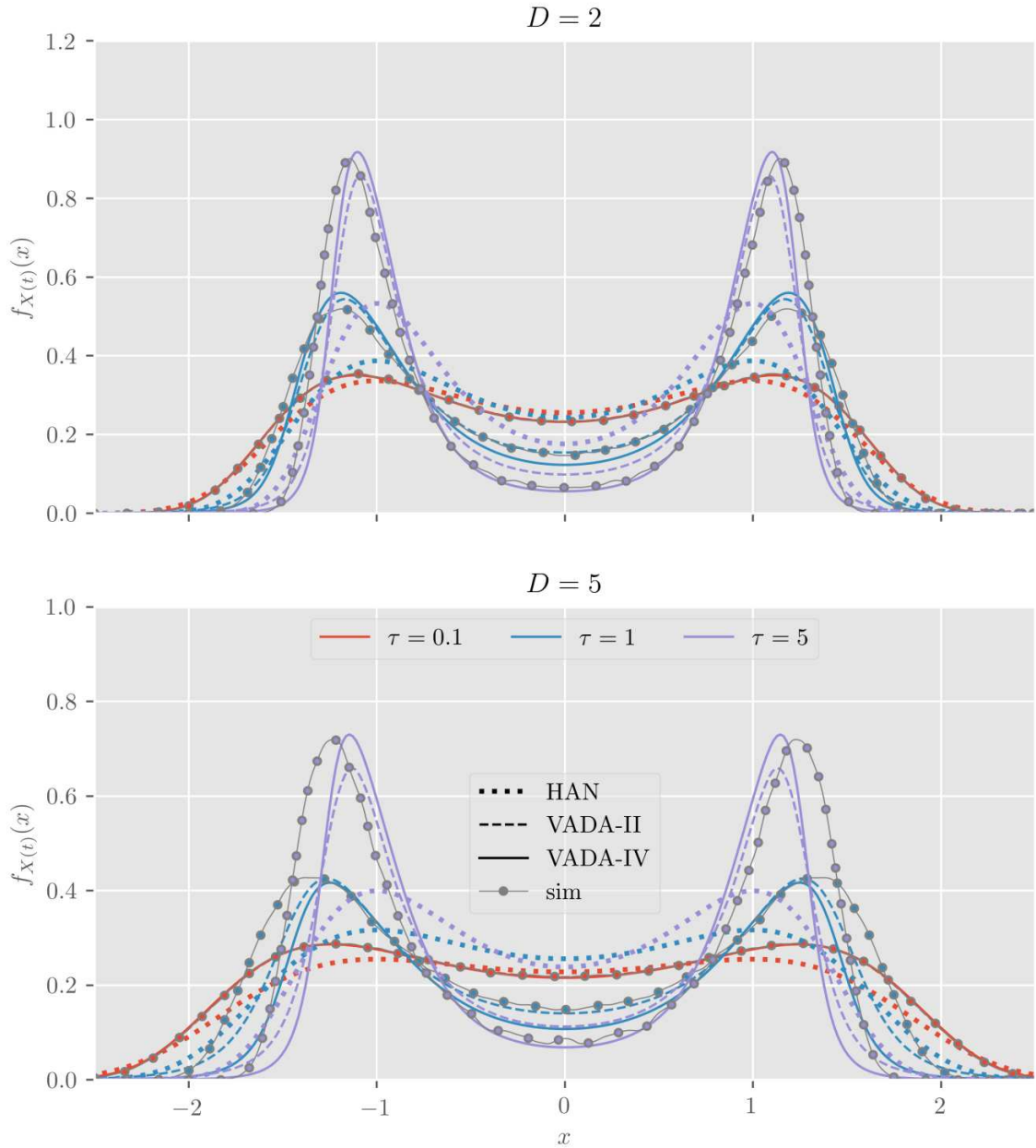

**Figure 4b.** Stationary response pdf for RDE (D1a,b) excited by a zero-mean OU process with $D = 2$ (upper panel) and $D = 5$ (lower panel) and for $\tau = 0.1, 1.0, 5.0$. Results from HAN, VADA-II and IV genFPK equations are presented along with MC simulations.

**Remark D3: On the prediction of peak value drift.** In Figs. 4a,b, we observe that, for large values of $D$ and $\tau$, the pdfs obtained by MC simulations exhibit peak values at abscissae larger than 1 in absolute value. This peak value drift, which has been documented before, see e.g. (Hänggi and Jung, 1995) p. 294, is predicted quite accurately by VADAs in the regime of $(D, \tau) = [0, 5.0] \times [0, 5.0]$, with VADA-IV being consistently more accurate than VADA-II. Note that this phenomenon is not captured at all by HAN, which predicts pdfs with peaks fixed at $\pm 1$.